\documentclass[preprint2]{aastex}

\shorttitle{}
\shortauthors{Richer et al.}

\begin{document}

\title{The Evolution of the Kinematics of Nebular Shells in \\ Planetary Nebulae in the Milky Way Bulge\footnote{The observations reported herein were acquired at the Observatorio Astron\'omico Nacional in the Sierra San Pedro M\'artir (OAN-SPM), B. C., Mexico.}}

\author{Michael G. Richer, Jos\'e Alberto L\'opez, Mar\'\i a Teresa Garc\'\i a-D\'\i az, \\ David M. Clark, Margarita Pereyra}
\affil{OAN, Instituto de Astronom\'\i a, Universidad Nacional Aut\'onoma de M\'exico, \\ P.O. Box 439027, San Diego, CA 92143}
\email{\{richer, jal, tere, dmclark, mally\}@astrosen.unam.mx}

\and

\author{Enrique D\'\i az-M\'endez}
\affil{Dept. of Physics and Astronomy, Texas Christian University, Fort Worth, Texas, U.S.A.  76109}
\email{e.d.mendez@tcu.edu}

\begin{abstract}

We study the line widths in the [\ion{O}{3}]$\lambda$5007 and H$\alpha$ lines for two groups of planetary nebulae in the Milky Way bulge based upon spectroscopy obtained at the \facility{Observatorio Astron\'omico Nacional in the Sierra San Pedro M\'artir (OAN-SPM)} using the Manchester Echelle Spectrograph.  The first sample includes objects early in their evolution, having high H$\beta$ luminosities, but [\ion{O}{3}]$\lambda 5007/\mathrm H\beta < 3$.  The second sample comprises objects late in their evolution, with \ion{He}{2}~$\lambda 4686/\mathrm H\beta > 0.5$.  These planetary nebulae represent evolutionary phases preceeding and following those of the objects studied by \citet{richeretal2008}.  Our sample of planetary nebulae with weak [\ion{O}{3}]$\lambda$5007 has a line width distribution similar to that of the expansion velocities of the envelopes of AGB stars, and shifted to systematically lower values as compared to the less evolved objects studied by \citet{richeretal2008}.  The sample with strong \ion{He}{2}~$\lambda 4686$ has a line width distribution indistinguishable from that of the more evolved objects from \citet{richeretal2008}, but a distribution in angular size that is systematically larger and so they are clearly more evolved.  These data and those of \citet{richeretal2008} form a homogeneous sample from a single Galactic population of planetary nebulae, from the earliest evolutionary stages until the cessation of nuclear burning in the central star.  They confirm the long-standing predictions of hydrodynamical models of planetary nebulae, where the kinematics of the nebular shell are driven by the evolution of the central star.   

\end{abstract}


\section{Introduction}

Hydrodynamical models of planetary nebulae have long predicted a particular kinematic evolution for the nebular shells, driven primarily by the evolution of the central stars \citep[e.g.,][]{kwoketal1978, kahnwest1985, schmidtvoigtkoppen1987a, schmidtvoigtkoppen1987b, breitschwerdtkahn1990, kahnbreitschwerdt1990, martenschonberner1991, mellema1994, villaveretal2002, perinottoetal2004, schonberneretal2007}.  Initially, the central stars are cool and their winds relatively slow.  This wind interacts with the wind that the precursor asymptotic giant branch (AGB) star emitted in a momentum-conserving mode \citep{kwok1982}.  However, the central star's temperature and wind velocity increase rapidly, with the consequences that an ionization front is driven through the AGB envelope and the interaction between the two winds switches to an energy driven mode, and a hot bubble is created behind the shocked wind.  The ionization front first accelerates the AGB envelope, now seen as the rim of the planetary nebula.  In time, once the internal pressure of the hot bubble exceeds that of the nebular shell, it further accelerates the nebular shell.  Theoretically, the latest phases of evolution are less clear, though the central star will cease nuclear energy generation, fade rapidly, and emit an ever-weaker wind, in principle allowing the inner part of the nebular envelope to backfill into the central cavity \citep[e.g.,][]{garciaseguraetal2006}.  

Although the many extant observations of the kinematics of planetary nebulae all show expansion of the nebular shell, there are few systematic observations of how these shells acquire their motion and how it evolves with time.  \citet{dopitaetal1985, dopitaetal1988} were the first to provide observational support for the early acceleration of the nebular shell from studies of planetary nebulae in the Magellanic Clouds.  Studies of Milky Way planetary nebulae provided much less convincing results \citep[e.g.,][]{chuetal1984, heap1993, medinaetal2006}.  Recently, \citet{richeretal2008} demonstrated the acceleration of nebular shells in bright planetary nebulae in the Milky Way bulge during the early evolution of their central stars and were able to associate the acceleration seen in different evolutionary stages with the phases of acceleration expected from theoretical models.  %

Here, we undertake a study that complements \citet{richeretal2008}, selecting objects earlier and later in their evolution than they did.  
The addition of these objects %
allows us to study the evolution of the kinematics of the nebular shell from the earliest stages of the planetary nebula phase until the cessation of nuclear burning in the central stars.  
In section \ref{sec_observations}, we present our new data and their analysis.  In section \ref{sec_results}, our results and their implications are outlined, the principal ones being the similarity of the line widths in H$\alpha$ and [\ion{O}{3}] $\lambda 5007$, that our sample of least evolved objects has a line width distribution shifted to the lowest values while the sample of most evolved objects has a size distribution with the largest sizes, and that the evolutionary state correlates with the H$\beta$ luminosity.  %
In section \ref{sec_conclusions} we summarize our conclusions.

\section{Observations and Analysis}\label{sec_observations}

\subsection{The Planetary Nebula Sample}\label{sec_sample_def}

Table \ref{table_objects} lists our two new samples of Bulge planetary nebulae.  There are 24 objects in the sample with low [\ion{O}{3}]$\lambda 5007/\mathrm H\beta$ ratios, all drawn from existing spectroscopic surveys \citep{allerkeyes1987, webster1988, cuisinieretal1996, cuisinieretal2000, ratagetal1997, escuderocosta2001, escuderoetal2004, exteretal2004, gornyetal2004}.  There are 21 objects in the sample with strong \ion{He}{2} 4686, only some of which have extensive spectroscopy (see the previous references).  Many of these objects were selected on the basis of the \ion{He}{2} 4686 line intensity from \citet{tylendaetal1994}.  Figure \ref{fig_o3_hb} explains the logic of the selection criteria for the two samples.  

The sample with weak [\ion{O}{3}] $\lambda 5007$ comprises objects entirely excluded from \citet{richeretal2008} and its selection were (i) a position within $10^{\circ}$ of the galactic centre, (ii) a large observed, reddening-corrected H$\beta$ flux, nominally $\log I(\mathrm H\beta) > -12.0$\,dex, and (iii) an intensity ratio [\ion{O}{3}]$\lambda 5007/\mathrm H\beta < 3$.  Since even the less evolved objects from \citet{richeretal2008} show the effects of acceleration of the nebular shell due to the passage of the ionization front, our last criterion should select objects that are significantly less evolved and thus probe the nebular kinematics at a stage that most closely reflects the kinematics of the undisturbed AGB envelope.  The weak [\ion{O}{3}] $\lambda 5007$ sample should include bright, young planetary nebulae whose central stars are still relatively cool.  All of the central stars should be on the horizontal part of their post-AGB evolutionary track.

The selection criteria for the \lq\lq strong \ion{He}{2}~$\lambda 4686$" sample were (i) a position within $10^{\circ}$ of the galactic centre and (ii) an intensity ratio \ion{He}{2}~$\lambda 4686 / \mathrm H\beta > 0.5$.  There was no restriction on the H$\beta$ flux.  The second criterion should select a sample of objects biased to more advanced evolutionary phases than the more evolved objects included in \citet{richeretal2008}, since their requirement for their more evolved objects was equivalent to a ratio \ion{He}{2}~$\lambda 4686 / \mathrm H\beta > 0.09$.  
This sample should include planetary nebulae whose central stars are at or slightly before their maximum temperature or fading towards the white dwarf cooling track.

\subsection{Observations, Data Reduction, and Derived Parameters}\label{obs_reduc_pars}

We acquired the observations reported here and measured the derived properties in a very similar fashion to our previous observations \citep[][]{richeretal2008}.  More details of our analysis may be found in \citet{richeretal2008,richeretal2009}.  

We obtained high resolution spectra with the Manchester echelle spectrometer \citep[MES-SPM; ][]{meaburnetal1984, meaburnetal2003} on 2008 June 1-2 and 10-17 and 2008 September 1-11 at the Observatorio Astron\'omico Nacional in the Sierra San Pedro M\'artir, Baja California, Mexico (OAN-SPM).  %
A 150\,$\mu$m wide slit ($1\farcs9$ wide on the sky, $5^{\prime}$ long) yielded a spectral resolution equivalent to 11\,km/s (2.6\,pix FWHM) and a spatial sampling of 0\farcs6/pixel when coupled to a SITe $1024\times 1024$ CCD with 24\,$\mu$m pixels binned $2\times 2$.  We used a ThAr lamp for the wavelength calibration, which typically yielded an internal precision better than $\pm 1.0$\,km/s. 

We usually obtained a single deep spectrum in each of the [\ion{O}{3}] $\lambda 5007$ and H$\alpha$ filters of, at most, 30 minutes duration.  %
When possible, the exposure time for the H$\alpha$ and [\ion{O}{3}] $\lambda 5007$ spectra were chosen to achieve similar signal levels.  The slit was oriented north-south and, except for M 2-38, centered on the object.  All of the planetary nebulae we observed are resolved (see Table \ref{table_objects}).  

The spectra were reduced using the twodspec and specred packages of the Image Reduction and Analysis Facility\footnote{IRAF is distributed by the National Optical Astronomical Observatories, which is operated by the Associated Universities for Research in Astronomy, Inc., under contract to the National Science Foundation.} (IRAF), following \citet[][Appendix B]{masseyetal1992}.  We edited the spectra to remove cosmic rays and subtracted a nightly mean bias image.  We rectified the object spectra and calibrated them in wavelength using the ThAr spectra.  Finally, wavelength-calibrated, one-dimensional spectra were extracted for each object.  No flux calibration was performed.  

The single exception to the foregoing was M 2-38, for which two 30 minute H$\alpha$ spectra were obtained at different positions.  The two spectra were reduced as described above and the final one-dimensional spectra were summed to produce the final H$\alpha$ spectrum.

We analyzed the one-dimensional spectra using a locally-implemented software package \citep[INTENS;][]{mccalletal1985}. %
INTENS models the emission line with a sampled gaussian function and models the continuum as a straight line.  For the strong \ion{He}{2}~$\lambda 4686$ sample, the H$\alpha$ line was usually accompanied by the \ion{He}{2}\,$\lambda$6560 line.  In these cases, a fit was made simultaneously to both lines and the continuum, assuming that both lines had the same width.    

Table \ref{table_objects} presents the observed line widths (FWHM; full width at half the maximum intensity) for each object in both H$\alpha$ and [\ion{O}{3}] $\lambda 5007$.  The formal uncertainties from fitting a Gaussian function with INTENS (one sigma; Table \ref{table_objects}) increase as the line width increases, a result of the line shape departing more from the Gaussian form at larger line widths \citep[][]{richeretal2009}. 
To obtain an idea of the real uncertainties, we measured the FWHM of the H$\alpha$ line profiles directly (using implot/IRAF).  For $\mathrm{FWHM}<1.2$\AA, there is no systematic difference between the line width and the Gaussian fit, though there is a dispersion of approximately $\pm 5\%$ of the line width.  For larger line widths, the Gaussian fit systematically underestimates the line width, with the difference reaching $8-9\%$ of the line width at $\mathrm{FWHM}\sim 2$\AA.  
We derive the true, intrinsic profile ($\sigma_{true}$), resulting from the kinematics of the planetary nebula, by correcting the observed profile ($\sigma_{obs}$) for instrumental ($\sigma_{inst}$), thermal ($\sigma_{th}$), and fine structure ($\sigma_{fs}$) broadening,

\begin{equation}
\sigma^2_{obs} = \sigma^2_{true} + \sigma^2_{inst} + \sigma^2_{th} + \sigma^2_{fs}\ .
\end{equation}

\noindent  We adopted an instrumental profile of FWHM of 2.6 pixels (measured: 2.5-2.7 pixels) for all objects.  We use the usual formula \citep[][eq. 2-243]{lang1980} to compute the thermal broadening, adopting rest wavelengths of 6562.83\AA\ and 5006.85\AA\ for H$\alpha$ and [\ion{O}{3}] $\lambda 5007$, respectively, assuming no turbulent velocity, and adopting the observed electron temperature, when available (preferably from [\ion{O}{3}] lines, but [\ion{N}{2}] otherwise).  If no electron temperature was available, we used the mean temperature for the other objects in each sample (weak [\ion{O}{3}] $\lambda 5007$ or high \ion{He}{2}\,4686).  %
For the fine structure broadening \citep{meaburn1970}, we adopted $\sigma_{fs} = 3.199$\,km/s for H$\alpha$ and zero for [\ion{O}{3}] $\lambda 5007$ \citep{garciadiazetal2008a}.

For real, spatially-resolved objects, the resulting line width, $\Delta V$, 

\begin{equation}
\Delta V = 2.3556\sigma_{true} 
\end{equation}

\noindent %
will be a luminosity-weighted, projected velocity width for the mass of the zone containing the emitting ion (O$^{2+}$ or H$^+$) enclosed within the spectrograph slit.  %
We adopt half of this intrinsic line width in velocity units %

\begin{equation}
\Delta V_{0.5} = 0.5\Delta V = 1.1778\sigma_{true}\ ,
\end{equation}  

\noindent %
as our measure of the kinematics for each object.

We obtain angular diameters (see Table \ref{table_objects}) by collapsing the spectra along the wavelength axis to produce one-dimensional H$\alpha$ spatial profiles, and then measured the diameter at 10\% of the peak intensity \citep{richeretal2008}.  The uncertainty in the diameters is less than half a pixel (0\farcs3) for the sample with weak [\ion{O}{3}]$\lambda$5007 and no more than a full pixel (0\farcs6) for the sample with strong \ion{He}{2}\,4686 (the H$\alpha$ profiles contained at least 320,000 and 71,000 photons, respectively).

\section{Results and Discussion}\label{sec_results}

We separate the Bulge planetary nebulae studied here and by \citet{richeretal2008} into four evolutionary groups based upon the properties of the central star.  The sample with weak [\ion{O}{3}] $\lambda 5007$ presented here has central stars whose temperatures are sufficiently low that only low ionization ions exist and so should be the least evolved.  Then follow the younger [\ion{O}{3}] $\lambda 5007$-bright objects from \citet[][\ion{He}{2}~$\lambda 6560$ absent]{richeretal2008}, with central stars that are hot enough to ionize O$^+$, but not O$^{2+}$ or He$^+$.  The central stars in both groups should be on the horizontal portion of their post-AGB evolutionary track in the H-R diagram.  The evolved [\ion{O}{3}] $\lambda 5007$-bright objects from \citet[][\ion{He}{2}~$\lambda 6560$ present]{richeretal2008} and our sample with strong \ion{He}{2}~$\lambda 4686$ have the hottest, most evolved central stars, with He$^{2+}$ present to differing degrees.  %
While we expect some overlap between the two groups, the strong \ion{He}{2}~$\lambda 4686$ should be biased to later evolutionary stages (\S \ref{sec_sample_def}).  These hotter central stars, particularly those with strong \ion{He}{2}~$\lambda 4686$, may have extinguished nuclear reactions and be fading towards the white dwarf domain.  

Had we assigned our planetary nebulae excitation classes instead of defining our four groups \citep[e.g.,][]{aller1956, gurzadyan1988, dopitaetal1990, reidparker2010}, there would be little difference in practice.  Our weak [\ion{O}{3}] $\lambda 5007$, young and evolved [\ion{O}{3}] $\lambda 5007$-bright, and strong \ion{He}{2}~$\lambda 4686$ groups correspond to very low, low-to-medium, medium-to-high, and high excitation classes, respectively.  Thus, our separation into evolutionary groups should be adequate for our purposes.

\subsection{H$\alpha$ and [\ion{O}{3}]$\lambda$5007 line widths}

We present the relation between the line widths ($\Delta V_{0.5}$) in [\ion{O}{3}]$\lambda$5007 and H$\alpha$ in Fig. \ref{fig_dvo3_dvha}.  As has been found previously for bright Bulge planetary nebulae, there is a near equality between these line widths \citep{richeretal2009, richeretal2010}.  The weak [\ion{O}{3}] $\lambda 5007$ data set extends this relation to smaller line widths while the sample with strong \ion{He}{2}~$\lambda 4686$ is well-mixed with previous data.  %

The tendency of finding systematically smaller line widths in [\ion{O}{3}]$\lambda$5007 at the smallest H$\alpha$ line widths in Fig. \ref{fig_dvo3_dvha} may be due to ionization stratification \citep[e.g.,][]{wilson1950}.  Hydrodynamical models clearly predict that the innermost parts of the ionized shell expand more slowly than the majority of the matter during the earliest phases of a planetary nebula's evolution \citep{villaveretal2002, perinottoetal2004} since 
the central star's wind is not fast enough to create a hot bubble.

The other feature in Fig. \ref{fig_dvo3_dvha} is the kink near H$\alpha$ line widths of about 33\,km/s.  
The objects that lie at higher velocities than the kink all have hot, evolved central stars.  
Presumably, the kink results from a drop in the projected outflow velocity in [\ion{O}{3}]$\lambda$5007, since that would seem energetically more feasible near the time when the central star is running out of nuclear energy.  Might the kink be the result of the loss of pressure from the central star's wind after nuclear reactions cease \citep[collapse of the hot bubble; e.g.,][]{garciaseguraetal2006, garciadiazetal2008b}?  Alternatively, the kink might also be produced via the passage of a \ion{He}{2} ionization front \citep[Figs. 3 and 5, ][]{schonberneretal2005}.  Hydrodynamical models do not clearly predict this behavior \citep{villaveretal2002, perinottoetal2004, schonberneretal2005, schonberneretal2007}.  %
More observations and detailed modeling of individual objects will be required to understand the origin of the kink.

\subsection{Evolution of the Nebular Kinematics}

Fig. \ref{fig_hist_lw} presents the line width distributions for the four groups of Bulge planetary nebulae.  The sample with weak [\ion{O}{3}] $\lambda 5007$ clearly differentiates itself from the young [\ion{O}{3}] $\lambda 5007$-bright objects, which is the evolutionary phase following it.  On the other hand, there is no noticeable difference between the kinematics of the evolved [\ion{O}{3}] $\lambda 5007$-bright objects and those with strong \ion{He}{2}~$\lambda 4686$.  

Statistical tests bear out these visual impressions.  %
The non-parametric U-test \citep[e.g.,][\S 5.4.3]{walljenkins2003} indicates that the probability of obtaining the line width distributions for the objects with weak [\ion{O}{3}] $\lambda 5007$ and the young [\ion{O}{3}] $\lambda 5007$-bright objects from the same parent distribution is only $1.0\times 10^{-4}$ when M3-13 is included (the highest line width in the weak [\ion{O}{3}] $\lambda 5007$ sample) or $1.4\times 10^{-5}$ if it is excluded.  Thus, the line width distributions for these two samples of objects are clearly different statistically.  Not surprisingly, there is no statistical evidence for any difference in the line width distributions for the evolved [\ion{O}{3}] $\lambda 5007$-bright objects and those with strong \ion{He}{2}~$\lambda 4686$.  

The line width distribution for the sample with weak [\ion{O}{3}] $\lambda 5007$ (Fig. \ref{fig_hist_lw}) is very similar to the distribution of envelope expansion velocities in AGB stars \citep{lewis1991, ramstedtetal2006}.  This is exactly what hydrodynamical models suggest for the earliest stages of evolution of the nebular shell before the central star's wind has created a hot bubble \citep{villaveretal2002, perinottoetal2004, schonberneretal2005, schonberneretal2007}.  Thus, this group of objects contains planetary nebulae whose nebular shells have not yet been significantly accelerated by the passage of the ionization front and stellar winds.  \citet{richeretal2008} have already argued that the young and evolved samples of [\ion{O}{3}] $\lambda 5007$-bright objects have the line width distributions expected from hydrodynamical models if they correspond, respectively, to the phase when a well-developed ionization front has swept through the nebular shell and the phase when  the hot bubble is actively accelerating the nebular shell.  

The line width distributions for the two groups of most evolved planetary nebulae
are indistinguishable. %
This implies that most of the nebular mass is not dramatically decelerated as the central star's %
luminosity decreases near the extinction of nuclear reactions, but continues expanding in a momentum-conserving mode.  %

\subsection{Nebular diameters}

Fig. \ref{fig_hist_d10} presents the distributions of angular diameters for the weak [\ion{O}{3}] $\lambda 5007$ and strong \ion{He}{2}~$\lambda 4686$ samples as well as those studied by \citet{richeretal2008}.  Although the sample with weak [\ion{O}{3}] $\lambda 5007$ are the least evolved, it does not have the smallest size distribution.  Its distribution of angular diameters is statistically different from that for the young [\ion{O}{3}] $\lambda 5007$-bright objects, with the U-test indicating a probability of only $6.1\times 10^{-4}$ of drawing the two distributions from the same parent distribution.  %

On the other hand, the distribution of diameters for the strong \ion{He}{2}~$\lambda 4686$ sample is clearly shifted to larger sizes compared to that for the evolved [\ion{O}{3}] $\lambda 5007$-bright objects.  The U-test indicates that the probability of drawing the two distributions from the same parent population is $1.3\times 10^{-6}$.  Assuming that the two groups are at the same distance, the difference in the distributions of angular sizes also argues that the objects with \ion{He}{2}~$\lambda 4686$ are the most evolved.  

Hydrodynamical models might help explain these results.  Models predict maximum H$\beta$ luminosities when the [\ion{O}{3}]$\lambda$5007 emission is strong \citep[e.g.,][]{schonberneretal2007}.  Our requirement that the weak [\ion{O}{3}]$\lambda$5007 objects have high luminosity could cause us to preferentially select objects closer than the Bulge, which would make them appear larger.  Also, models predict important structural changes during the early evolution, depending upon the relative importance of the ionization front and hot bubble \citep[e.g.,][]{perinottoetal2004} that could affect the sizes we measure.  Whatever the reason, a lack of a strong correlation between nebular diameter and evolutionary indicators have been obtained before \citep{chuetal1984,gurzadyan1997}.

\subsection{H$\beta$ luminosities}

In Fig. \ref{fig_dv_hb}, we plot the line width as a function of H$\beta$ luminosity for the four planetary nebula samples.   Although the H$\beta$ luminosities are not very accurate \citep[][]{richeretal2008}, there is a very clear progression as a function of evolutionary state, from the weak [\ion{O}{3}] $\lambda 5007$ sample, with the smallest line widths and high luminosities, through the young and evolved [\ion{O}{3}] $\lambda 5007$-bright objects to the strong \ion{He}{2}~$\lambda 4686$ sample, which has large line widths and the faintest luminosities.  

\begin{figure}
\begin{center}
\includegraphics[width=\columnwidth,angle=0]{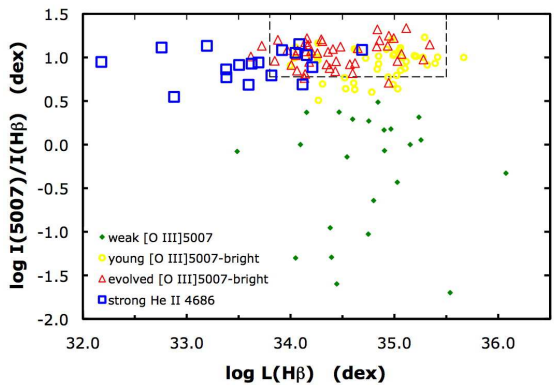}
\end{center}
\caption{The selection criteria for the two samples of Bulge planetary nebulae presented here were designed to select planetary nebulae that are (a) less evolved and (b) more evolved than the sample studied by \citet{richeretal2008}.  
The area within the dotted box is approximately the area occupied by bright extragalactic planetary in stellar systems without ongoing star formation %
\citep{jacobyciardullo1999, richeretal1999, walshetal1999, dudziaketal2000, rothetal2004, mendezetal2005, zijlstraaetal2006, goncalvesetal2007, richermccall2008}.  \label{fig_o3_hb}}
\end{figure}

The mixing of objects from the different samples in Figs. \ref{fig_o3_hb} and \ref{fig_dvo3_dvha}-\ref{fig_dv_hb} is undoubtedly due to the different effects that can affect the expansion of the nebular shell.  Although the range in masses of the progenitor stars is likely relatively small, given their ages, the range in metallicity is substantially larger \citep[e.g.,][]{sahuetal2006, zoccalietal2008}.  At lower metallicity the envelope expansion velocities for AGB stars are also lower \citep[e.g.,][]{woodetal1992, marshalletal2004, mattssonetal2008, wachteretal2008, groenewegenetal2009}.  On the other hand, hydrodynamical models %
find larger accelerations of the nebular shell at lower metallicity due to the higher electron temperature \citep{schonberneretal2005b}.  %
The progenitor mass should affect the %
nebular kinematics primarily via the central star mass %
and its strong influence upon the wind output and evolutionary time scale %
\citep[e.g.,][]{villaveretal2002, perinottoetal2004}.  %
Considering that the nebular shells are not likely to always be spherical, projection effects will further affect the measured line widths.  Thus, the dispersion seen in Figs. \ref{fig_o3_hb} and \ref{fig_dvo3_dvha}-\ref{fig_dv_hb} would seem plausible from natural causes.

\section{Conclusions}\label{sec_conclusions}

We have obtained kinematic data for two samples of planetary nebulae in the Milky Way bulge, selected so as to include objects very early and late in their evolution (\S \ref{sec_sample_def}).  %
We measure line widths for H$\alpha$ and [\ion{O}{3}] $\lambda 5007$ in most cases.  We combine these data sets with that studied by \citet{richeretal2008}.  We define four evolutionary groups, based upon the temperature of the central star, which allow us to study the kinematics of the nebular shell from the earliest phases until the central star ceases nuclear burning.  

Generally, we find a near equality of the line widths for the H$\alpha$ and [\ion{O}{3}] $\lambda 5007$ lines in any given object.
Ionization stratification likely accounts for the deviations: The [\ion{O}{3}] $\lambda 5007$ line widths are systematically smaller than the H$\alpha$ line widths for the smallest and largest H$\alpha$ line widths, corresponding to the earliest and latest evolutionary phases, respectively.

We see clear evolution of the kinematics of the nebular shell.  The least evolved objects, our planetary nebulae with weak [\ion{O}{3}] $\lambda 5007$, have cool central stars  and the nebular envelopes have a line width distribution similar to that of the envelope expansion velocities of AGB stars, indicating that the ionization front has not yet been able to substantially accelerate the nebular shell.  In subsequent phases \citep{richeretal2008}, the nebular shell is first accelerated by the passage of an ionization front and then further accelerated once the central star's wind produces a hot bubble.  The line width distributions for the planetary nebulae in these three evolutionary phases are statistically distinct.  The most evolved objects, with high \ion{He}{2}~$\lambda 4686$ ratios, have a similar line width distribution to evolved [\ion{O}{3}] $\lambda 5007$-bright objects, suggesting that no further acceleration occurs as the central stars reach their highest temperatures, their nuclear reactions cease, and their winds decline.  %

This kinematic evolution of the nebular shell has long been predicted by hydrodynamical models \citep[e.g.,][]{kahnwest1985, martenschonberner1991, mellema1994, villaveretal2002, perinottoetal2004}.  %
Our results, together with those of \citet{richeretal2008}, based upon a large sample of Galactic planetary nebulae from a single stellar population, clearly confirm these predictions.  At least until the point at which nuclear reactions cease in the central stars, their ionizing fluxes and winds continuously accelerate the nebular envelopes.  What happens thereafter is not yet completely clear, and would require samples of planetary nebulae chosen specifically to contain hot central stars of low luminosity.  %

\acknowledgments

We thank the technical personnel at the OAN-SPM, and particularly Gabriel Garc\'\i a, Gustavo Melgoza, Salvador Monrroy, and Felipe Montalvo who were the telescope operators during our observing runs.  Their excellent support was a great help in obtaining the data presented here.  We acknowledge financial support throughout this project from CONACyT through grants 43121, 49447, and 82066 and from UNAM-DGAPA via grants IN108406-2 and IN116908-3.  This research has made use of the SIMBAD database, operated at CDS, Strasbourg, France

\clearpage

\begin{deluxetable}{llllcccccc}
\tabletypesize{\scriptsize}
\rotate
\tablecaption{Bulge Planetary Nebula Sample\label{table_objects}} 
\tablewidth{0pt}
\tablehead{
\colhead{Object} & \colhead{PN G} & \colhead{sample} & \colhead{Run} & \colhead{FWHM(H$\alpha$)$^{\mathrm a}$} & \colhead{$\Delta V_{0.5}(\mathrm H\alpha)$} & \colhead{FWHM(5007)$^{\mathrm a}$} & \colhead{$\Delta V_{0.5}(5007)$} & \colhead{H$\alpha$ Diameter$^{\mathrm b}$} & \colhead{$L(\mathrm H\beta)$} \\
\colhead{} & \colhead{} & \colhead{} & \colhead{} & \colhead{(\AA)} & \colhead{(km/s)} & \colhead{(\AA)} & \colhead{(km/s)} & \colhead{10\% $I_{max}$} &\colhead{(erg/s)}
}
\startdata
H 1-24   & 004.6+06.0 & weak 5007    & 2008jun & $ 0.8783 \pm 0.0044 $ & $ 11.22 \pm 0.10 $ & $ 0.6288 \pm 0.0081 $ & $ 17.54 \pm 0.24 $ &  7.3 & 34.84 \\
H 1-34   & 005.5+02.7 & weak 5007    & 2008jun & $ 0.8168 \pm 0.0072 $ & $ 12.31 \pm 0.16 $ & $ 0.5422 \pm 0.0077 $ & $ 14.88 \pm 0.23 $ &  5.0 & 34.60 \\
H 1-39   & 356.5-03.9 & weak 5007    & 2008jun & $ 0.7737 \pm 0.0019 $ & $ 11.31 \pm 0.04 $ & $ 0.3343 \pm 0.0030 $ & $  7.77 \pm 0.09 $ &  5.9 & 34.80 \\
H 1-43   & 357.1-04.7 & weak 5007    & 2008jun & $ 0.7230 \pm 0.0030 $ & $ 10.82 \pm 0.07 $ & $                   $ & $                $ &  5.8 & 34.45 \\
H 1-44   & 358.9-03.7 & weak 5007    & 2008jun & $ 0.6198 \pm 0.0022 $ & $  5.37 \pm 0.05 $ & $ 0.2950 \pm 0.0028 $ & $  6.19 \pm 0.08 $ &  6.3 & 34.10 \\
H 1-55   & 001.7-04.4 & weak 5007    & 2008jun & $ 0.6369 \pm 0.0014 $ & $  6.30 \pm 0.03 $ & $ 0.2939 \pm 0.0047 $ & $  6.18 \pm 0.14 $ &  5.7 & 34.38 \\
H 2-25   & 004.8+02.0 & weak 5007    & 2008jun & $ 0.6923 \pm 0.0023 $ & $  2.37 \pm 0.05 $ & $ 0.2747 \pm 0.0087 $ & $  4.79 \pm 0.26 $ &  7.1 & 34.54 \\
H 2-29   & 357.6-03.3 & weak 5007    & 2008jun & $ 0.9433 \pm 0.0062 $ & $ 15.28 \pm 0.14 $ & $                   $ & $                $ & 12.7 & 33.49 \\
H 2-48   & 011.3-09.4 & weak 5007    & 2008jun & $ 0.7564 \pm 0.0012 $ & $  9.45 \pm 0.03 $ & $ 0.3293 \pm 0.0015 $ & $  7.42 \pm 0.04 $ &  5.9 & 35.26 \\
He 2-260 & 008.2+06.8 & weak 5007    & 2008jun & $ 0.7031 \pm 0.0012 $ & $  6.48 \pm 0.03 $ & $ 0.2903 \pm 0.0035 $ & $  5.91 \pm 0.10 $ &  7.7 & 34.39 \\
M 1-26   & 358.9-00.7 & weak 5007    & 2008jun & $ 0.7352 \pm 0.0023 $ & $ 10.13 \pm 0.05 $ & $ 0.3477 \pm 0.0031 $ & $  8.23 \pm 0.09 $ &  7.3 & 36.07 \\
M 1-27   & 356.5-02.3 & weak 5007    & 2008jun & $ 0.7270 \pm 0.0018 $ & $ 10.43 \pm 0.04 $ & $                   $ & $                $ &  8.9 & 35.54 \\
M 1-30   & 355.9-04.2 & weak 5007    & 2008jun & $ 0.8701 \pm 0.0029 $ & $ 14.10 \pm 0.07 $ & $ 0.5549 \pm 0.0058 $ & $ 15.32 \pm 0.17 $ &  6.5 & 34.97 \\
M 1-44   & 004.9-04.9 & weak 5007    & 2008jun & $ 0.6746 \pm 0.0030 $ & $  8.13 \pm 0.07 $ & $                   $ & $                $ &  7.6 & 34.75 \\
M 1-45   & 012.6-02.6 & weak 5007    & 2008jun & $ 0.7768 \pm 0.0028 $ & $ 12.15 \pm 0.06 $ & $                   $ & $                $ &  5.6 & 35.15 \\
M 2-10   & 354.2+04.3 & weak 5007    & 2008jun & $ 0.6916 \pm 0.0010 $ & $  9.00 \pm 0.02 $ & $ 0.4007 \pm 0.0024 $ & $ 10.21 \pm 0.07 $ &  7.5 & 34.90 \\
M 2-12   & 359.8+05.6 & weak 5007    & 2008jun & $ 0.6136 \pm 0.0011 $ & $  4.81 \pm 0.03 $ & $ 0.2955 \pm 0.0447 $ & $  6.13 \pm 1.34 $ &  7.1 & 34.56 \\
M 2-14   & 003.6+03.1 & weak 5007    & 2008jun & $ 0.7703 \pm 0.0038 $ & $ 11.30 \pm 0.09 $ & $ 0.4426 \pm 0.0048 $ & $ 11.63 \pm 0.14 $ &  5.3 & 34.47 \\
M 2-19   & 000.2-01.9 & weak 5007    & 2008jun & $ 0.7534 \pm 0.0022 $ & $ 10.59 \pm 0.05 $ & $ 0.3885 \pm 0.0036 $ & $  9.73 \pm 0.11 $ &  8.8 & 34.75 \\
M 2-7    & 353.7+06.3 & weak 5007    & 2008jun & $ 0.8499 \pm 0.0026 $ & $ 14.13 \pm 0.06 $ & $ 0.5337 \pm 0.0035 $ & $ 14.61 \pm 0.10 $ &  9.9 & 34.15 \\
M 3-13   & 005.2+04.2 & weak 5007    & 2008jun & $ 1.2876 \pm 0.0282 $ & $ 26.20 \pm 0.64 $ & $ 0.9708 \pm 0.0556 $ & $ 28.38 \pm 1.67 $ &  6.2 & 35.24 \\
M 3-17   & 359.3-03.1 & weak 5007    & 2008jun & $ 0.8316 \pm 0.0020 $ & $ 13.47 \pm 0.05 $ & $ 0.4159 \pm 0.0051 $ & $ 10.70 \pm 0.15 $ &  5.7 & 34.90 \\
SwSt 1   & 001.5-06.7 & weak 5007    & 2008jun & $ 0.9277 \pm 0.0027 $ & $ 16.43 \pm 0.06 $ & $ 0.5919 \pm 0.0033 $ & $ 16.54 \pm 0.10 $ &  4.6 & 35.03 \\
Th 3-16  & 357.5-03.1 & weak 5007    & 2008jun & $ 0.6639 \pm 0.0017 $ & $  7.10 \pm 0.04 $ & $                   $ & $                $ &  6.3 & 34.05 \\
Al 1     & 006.8-08.6 & strong 4686  & 2008sep & $ 1.2782 \pm 0.0056 $ & $ 24.04 \pm 0.17 $ & $ 0.7609 \pm 0.0162 $ & $ 21.73 \pm 0.49 $ & 13.0 & 34.12 \\
Al 2-E   & 359.3+03.6 & strong 4686  & 2008sep & $ 1.6281 \pm 0.0246 $ & $ 33.31 \pm 0.74 $ & $ 1.0817 \pm 0.0283 $ & $ 31.66 \pm 0.85 $ &  8.6 & 33.19 \\
Al 2-H   & 357.2+01.4 & strong 4686  & 2008sep & $ 1.0106 \pm 0.0065 $ & $ 16.07 \pm 0.19 $ & $ 0.7402 \pm 0.0180 $ & $ 21.08 \pm 0.54 $ &  9.3 &       \\
Al 2-I   & 359.5+02.6 & strong 4686  & 2008sep & $ 1.1852 \pm 0.0099 $ & $ 21.41 \pm 0.30 $ & $ 0.9063 \pm 0.0210 $ & $ 26.25 \pm 0.63 $ &  6.6 &       \\
H 2-44   & 005.5-04.0 & strong 4686  & 2008sep & $ 1.4812 \pm 0.0266 $ & $ 29.51 \pm 0.80 $ & $ 1.1545 \pm 0.0395 $ & $ 33.89 \pm 1.18 $ & 12.4 & 34.16 \\
KFL 02   &            & strong 4686  & 2008sep & $ 1.2658 \pm 0.0102 $ & $ 23.70 \pm 0.31 $ & $ 0.9265 \pm 0.0169 $ & $ 26.89 \pm 0.51 $ &  8.6 & 31.49 \\
KFL 09   &            & strong 4686  & 2008sep & $ 1.3491 \pm 0.0240 $ & $ 25.99 \pm 0.72 $ & $ 1.0010 \pm 0.0259 $ & $ 29.18 \pm 0.78 $ & 12.0 & 33.82 \\
KFL 16   & 005.6-04.7 & strong 4686  & 2008sep & $ 1.5910 \pm 0.0254 $ & $ 32.36 \pm 0.76 $ & $ 1.2402 \pm 0.0450 $ & $ 36.51 \pm 1.35 $ & 13.9 & 33.38 \\
M 2-38   & 005.7-05.3 & strong 4686  & 2008sep & $ 1.3917 \pm 0.0185 $ & $ 27.95 \pm 0.55 $ & $ 0.9841 \pm 0.0315 $ & $ 28.71 \pm 0.94 $ & 14.4 & 33.50 \\
M 3-22   & 000.7-03.7 & strong 4686  & 2008sep & $ 1.2684 \pm 0.0125 $ & $ 23.70 \pm 0.37 $ & $ 0.9609 \pm 0.0250 $ & $ 27.94 \pm 0.75 $ &  9.3 & 34.21 \\
M 3-23   & 000.9-04.8 & strong 4686  & 2008sep & $ 1.4720 \pm 0.0291 $ & $ 29.27 \pm 0.87 $ & $ 1.0804 \pm 0.0538 $ & $ 31.62 \pm 1.61 $ & 14.0 & 34.69 \\
Pe 1-12  & 004.0-05.8 & strong 4686  & 2008sep & $ 1.5601 \pm 0.0428 $ & $ 31.84 \pm 1.28 $ & $ 1.2239 \pm 0.0589 $ & $ 36.03 \pm 1.76 $ & 11.9 & 33.38 \\
Pe 1-13  & 010.7-06.7 & strong 4686  & 2008sep & $ 1.0811 \pm 0.0059 $ & $ 18.42 \pm 0.18 $ & $                   $ & $                $ &  9.4 & 33.69 \\
Pe 2-13  & 006.4-04.6 & strong 4686  & 2008sep & $ 1.3648 \pm 0.0129 $ & $ 26.42 \pm 0.39 $ & $                   $ & $                $ &  9.8 & 34.08 \\
Sa 2-230 & 010.7+07.4 & strong 4686  & 2008sep & $ 1.7239 \pm 0.0576 $ & $ 35.74 \pm 1.73 $ & $ 1.1654 \pm 0.0490 $ & $ 34.23 \pm 1.47 $ & 15.1 & 33.63 \\
SB 15    & 009.3-06.5 & strong 4686  & 2008sep & $ 1.2971 \pm 0.0067 $ & $ 23.40 \pm 0.20 $ & $ 0.9568 \pm 0.0116 $ & $ 27.76 \pm 0.35 $ &  8.5 & 32.76 \\
SB 37    & 352.6-04.9 & strong 4686  & 2008sep & $ 1.5842 \pm 0.0175 $ & $ 32.18 \pm 0.52 $ & $ 1.2002 \pm 0.0297 $ & $ 35.29 \pm 0.89 $ & 10.0 & 33.92 \\
SB 38    & 352.7-08.4 & strong 4686  & 2008sep & $ 1.4677 \pm 0.0154 $ & $ 28.59 \pm 0.46 $ & $ 1.1736 \pm 0.0255 $ & $ 34.43 \pm 0.76 $ & 14.0 & 32.18 \\
SB 55    & 359.4-08.5 & strong 4686  & 2008sep & $ 1.4107 \pm 0.0356 $ & $ 25.56 \pm 1.07 $ & $ 1.0770 \pm 0.0490 $ & $ 31.41 \pm 1.47 $ & 14.9 & 32.88 \\
ShWi 2-1 & 001.4-03.4 & strong 4686  & 2008sep & $ 1.0534 \pm 0.0250 $ & $ 17.45 \pm 0.75 $ & $                   $ & $                $ & 11.8 & 33.60 \\
Th 3-26  & 358.8+03.0 & strong 4686  & 2008sep & $ 1.0768 \pm 0.0090 $ & $ 20.97 \pm 0.27 $ & $ 0.6937 \pm 0.0103 $ & $ 19.78 \pm 0.31 $ &  9.7 & 34.05 \\
H 2-13   & 357.2+02.0 & 6560 present & 2008jun & $ 0.9376 \pm 0.0023 $ & $ 16.72 \pm 0.05 $ & $ 0.5785 \pm 0.0045 $ & $ 16.12 \pm 0.13 $ &  6.0 & 34.47 \\
H 2-43   & 003.4-04.8 & 6560 absennt & 2008jun & $ 1.9363 \pm 0.0185 $ & $ 39.20 \pm 0.42 $ & $ 1.7300 \pm 0.0658 $ & $ 51.28 \pm 1.97 $ &  4.6 & 34.88 \\
\enddata
\tablenotetext{a}{This is the observed line width, uncorrected for instrumental, thermal, or fine structure broadening.}
\tablenotetext{b}{The diameter was measured by collapsing the H$\alpha$ spectra into one-dimensional spatial profiles and measuring the diameter at 10\% of the maximum intensity.}
\end{deluxetable}

\clearpage

\begin{figure*}
\begin{center}
\includegraphics[height=\columnwidth,angle=90]{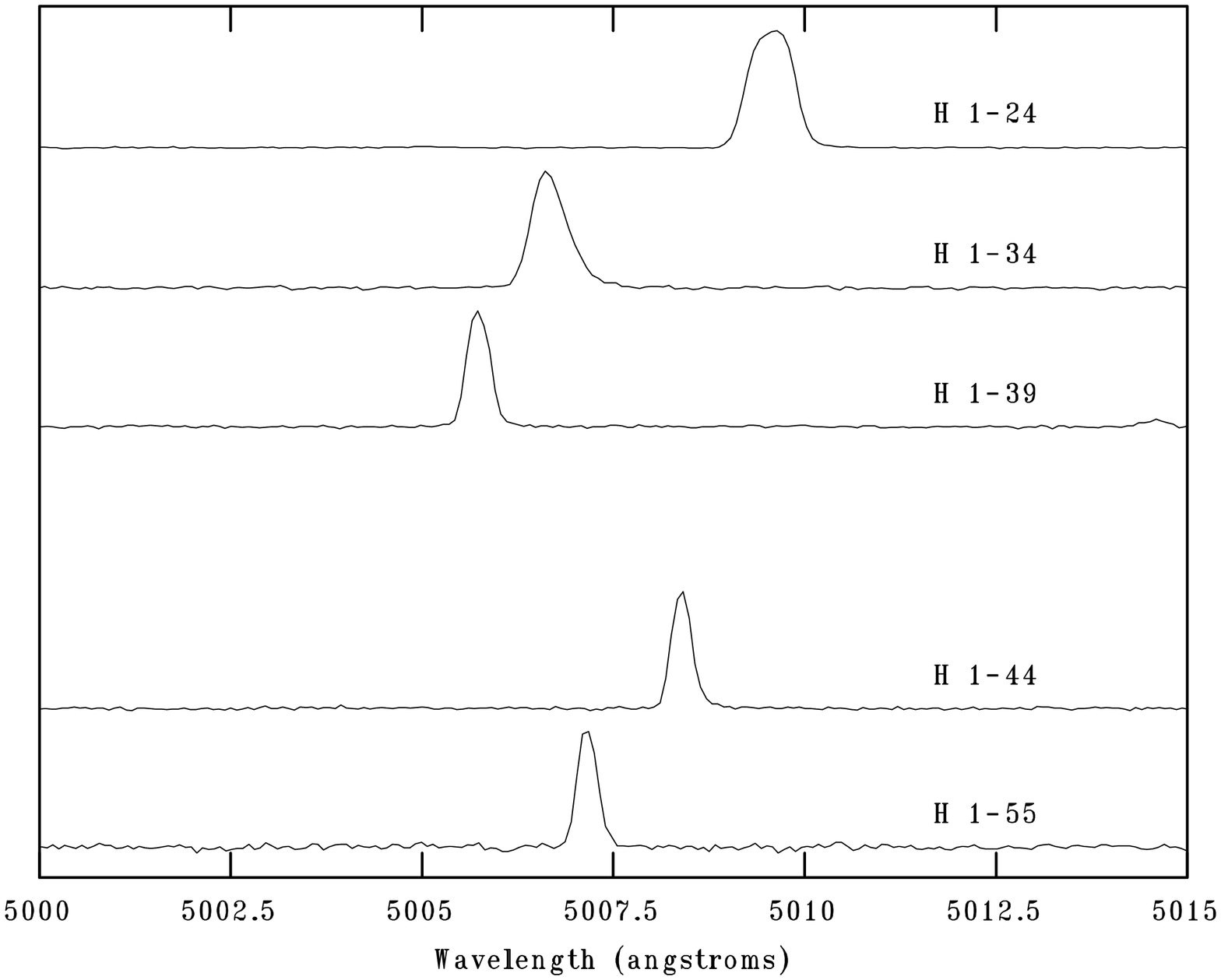}
\includegraphics[height=\columnwidth,angle=90]{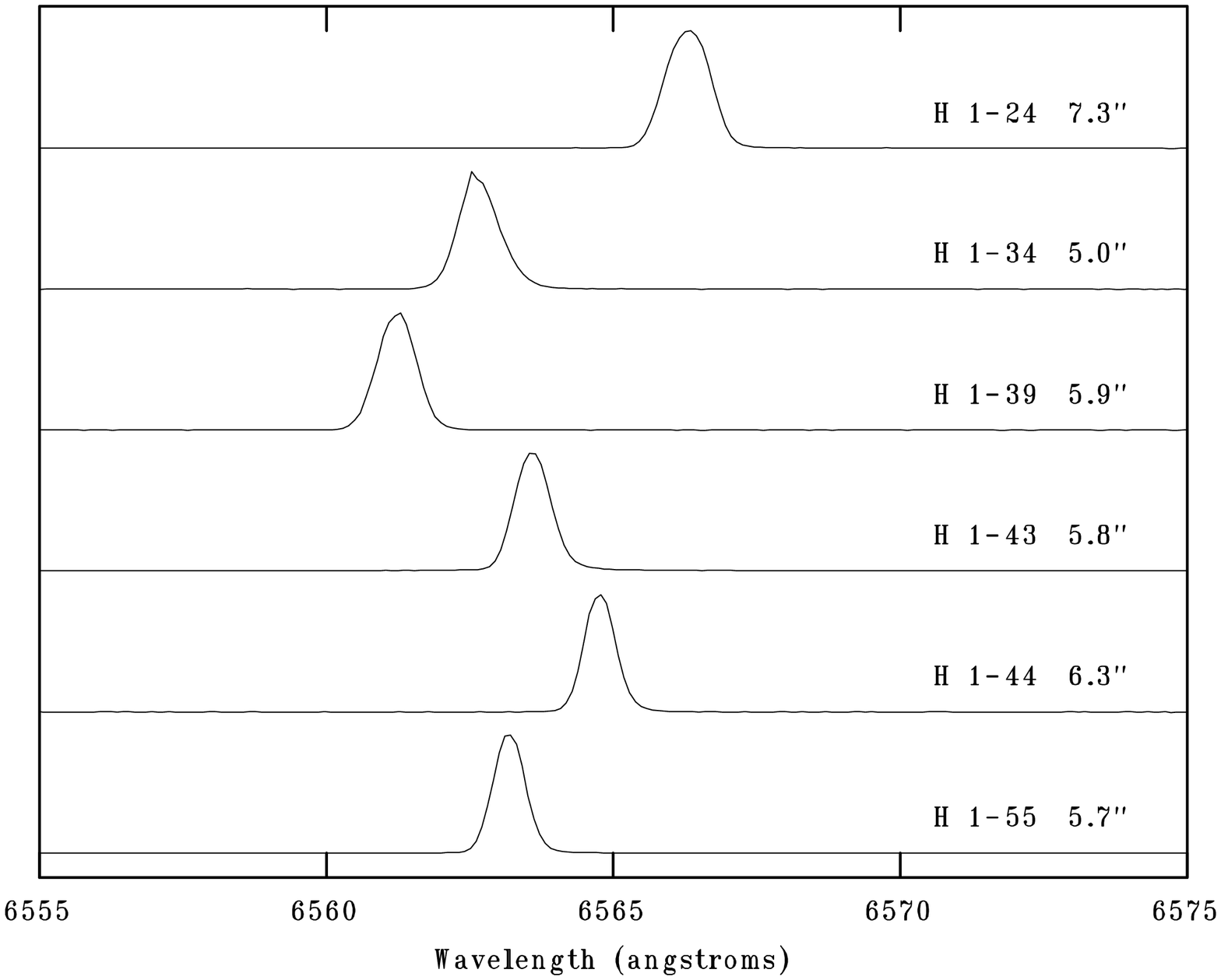} 
\includegraphics[height=\columnwidth,angle=90]{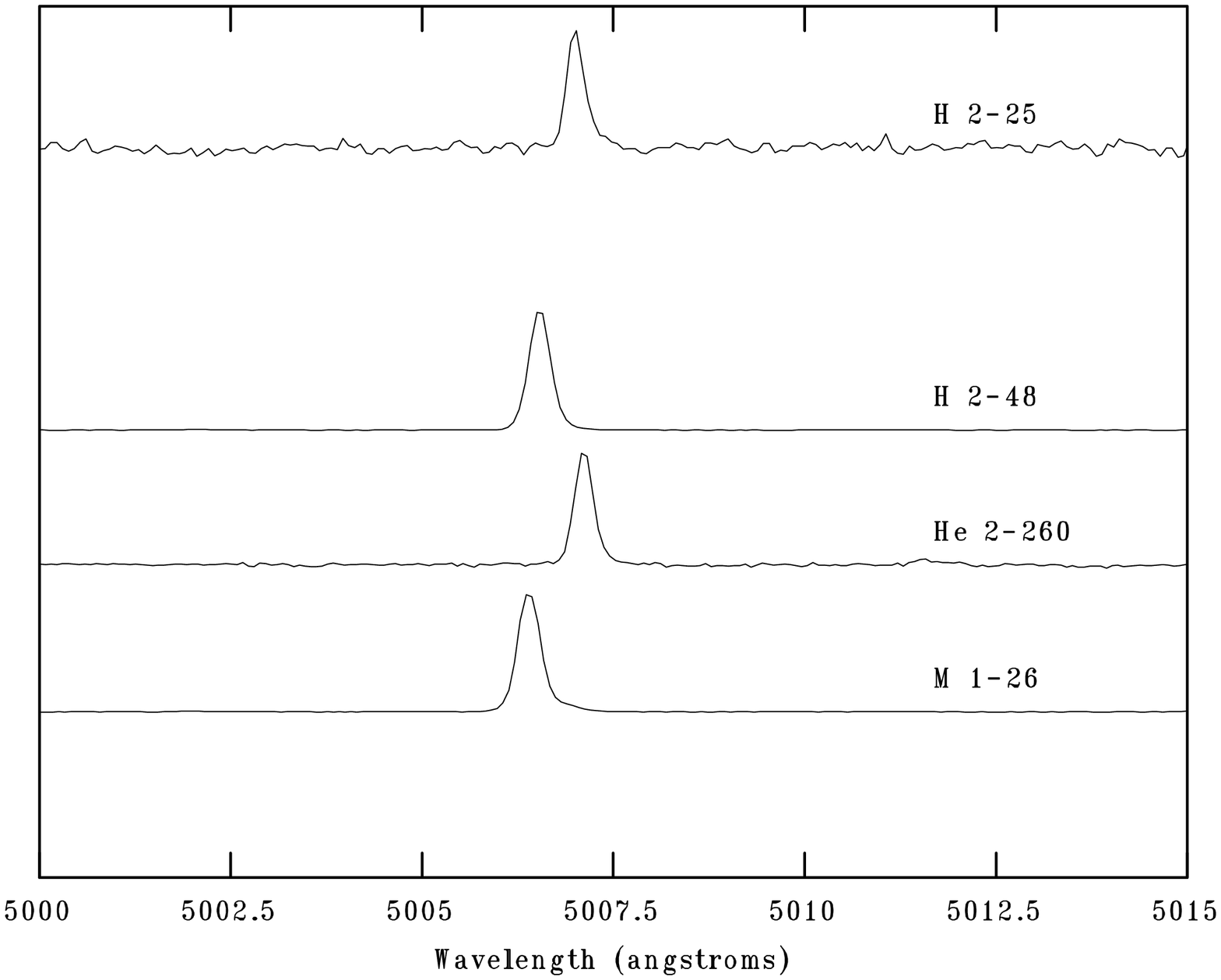}
\includegraphics[height=\columnwidth,angle=90]{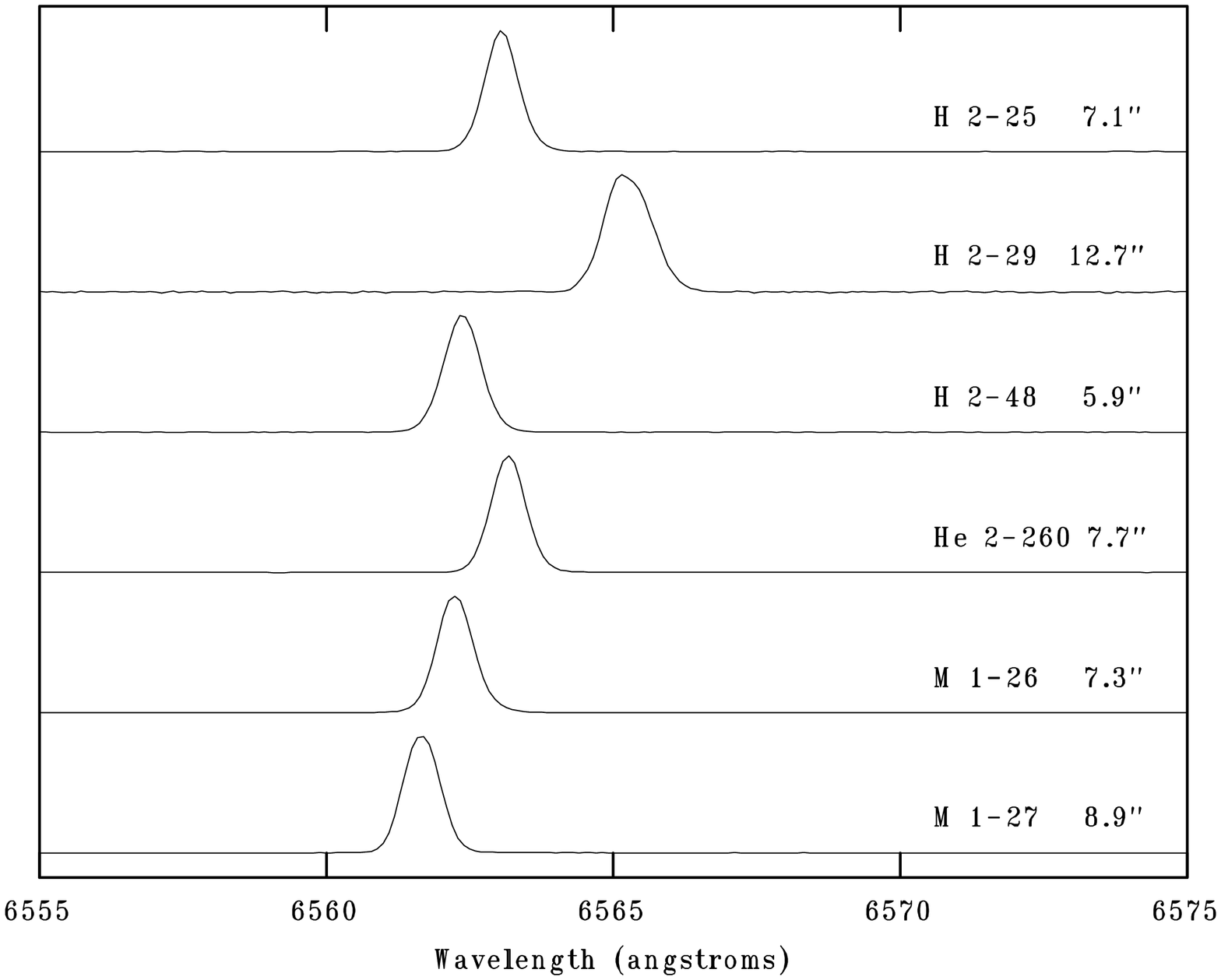}
\end{center}
\caption{We present the line profiles in [\ion{O}{3}]$\lambda$5007 and H$\alpha$ for the sample of planetary nebulae with weak [\ion{O}{3}]$\lambda$5007 lines.  The wavelength interval plotted is always that indicated on the bottom panels.  The number beside the name in the H$\alpha$ panels is the angular diameter of the object at 10\% of maximum intensity.    \label{fig_gallery_weak1}}
\end{figure*}

\begin{figure*}
\begin{center}
\includegraphics[height=\columnwidth,angle=90]{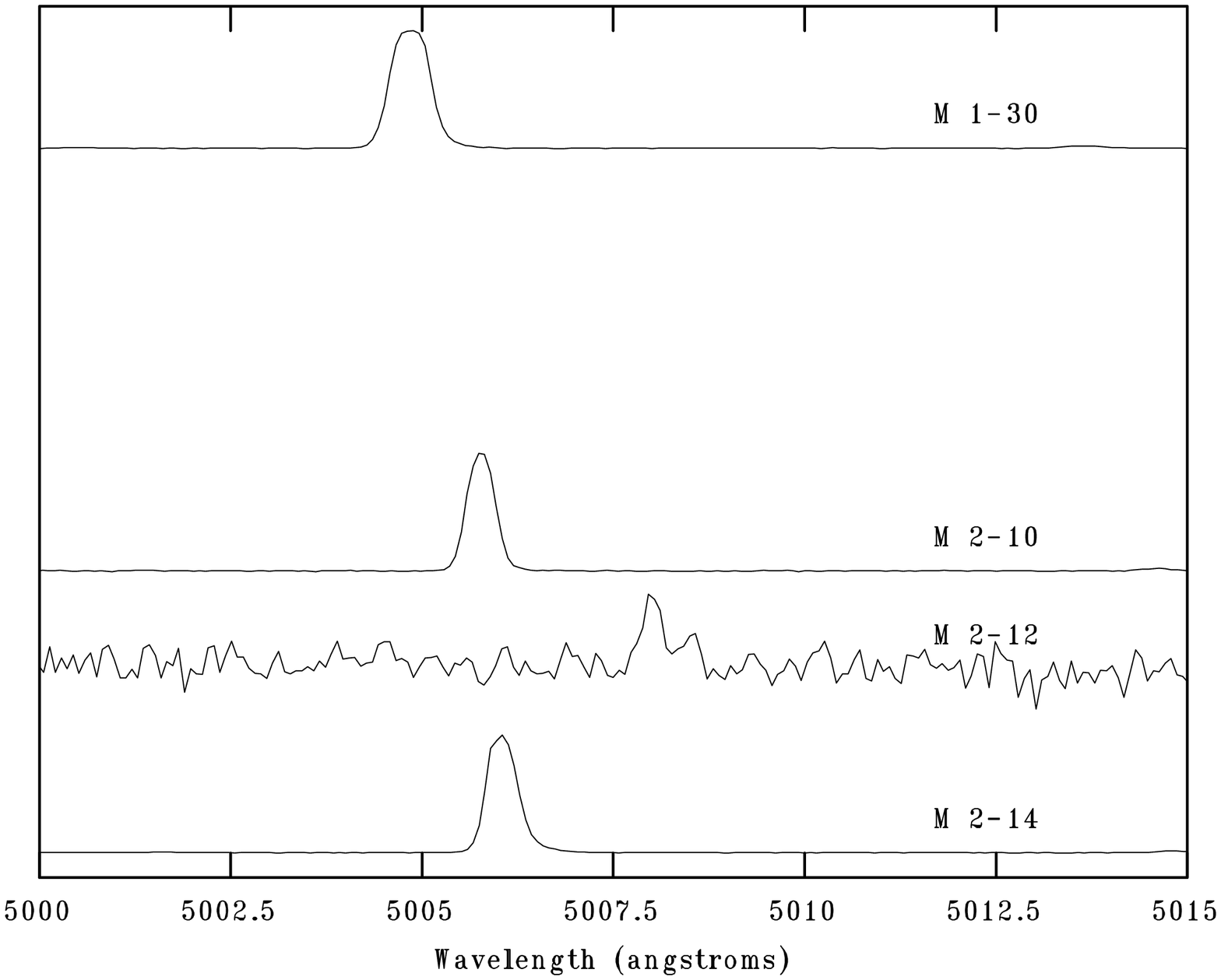}
\includegraphics[height=\columnwidth,angle=90]{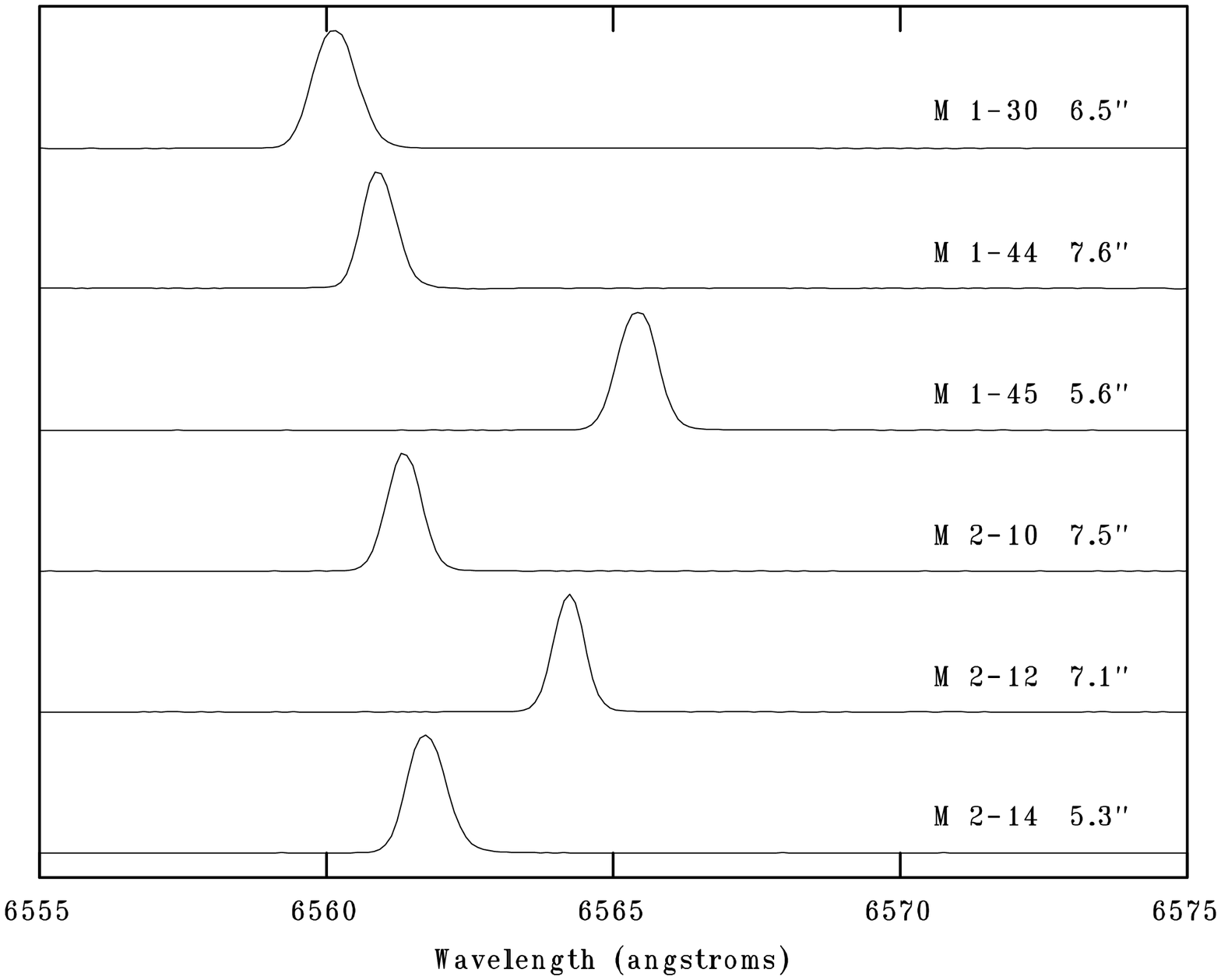} 
\includegraphics[height=\columnwidth,angle=90]{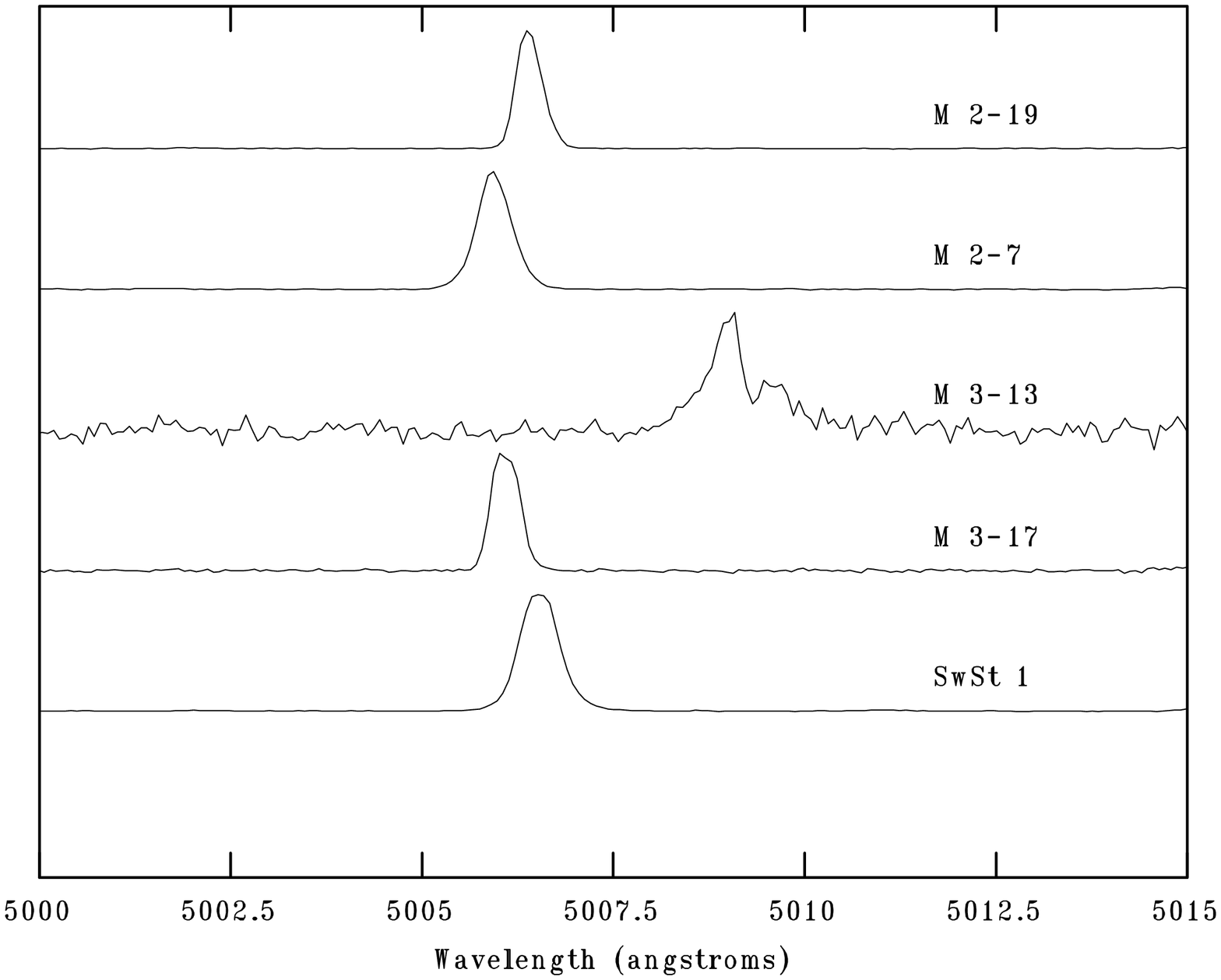}
\includegraphics[height=\columnwidth,angle=90]{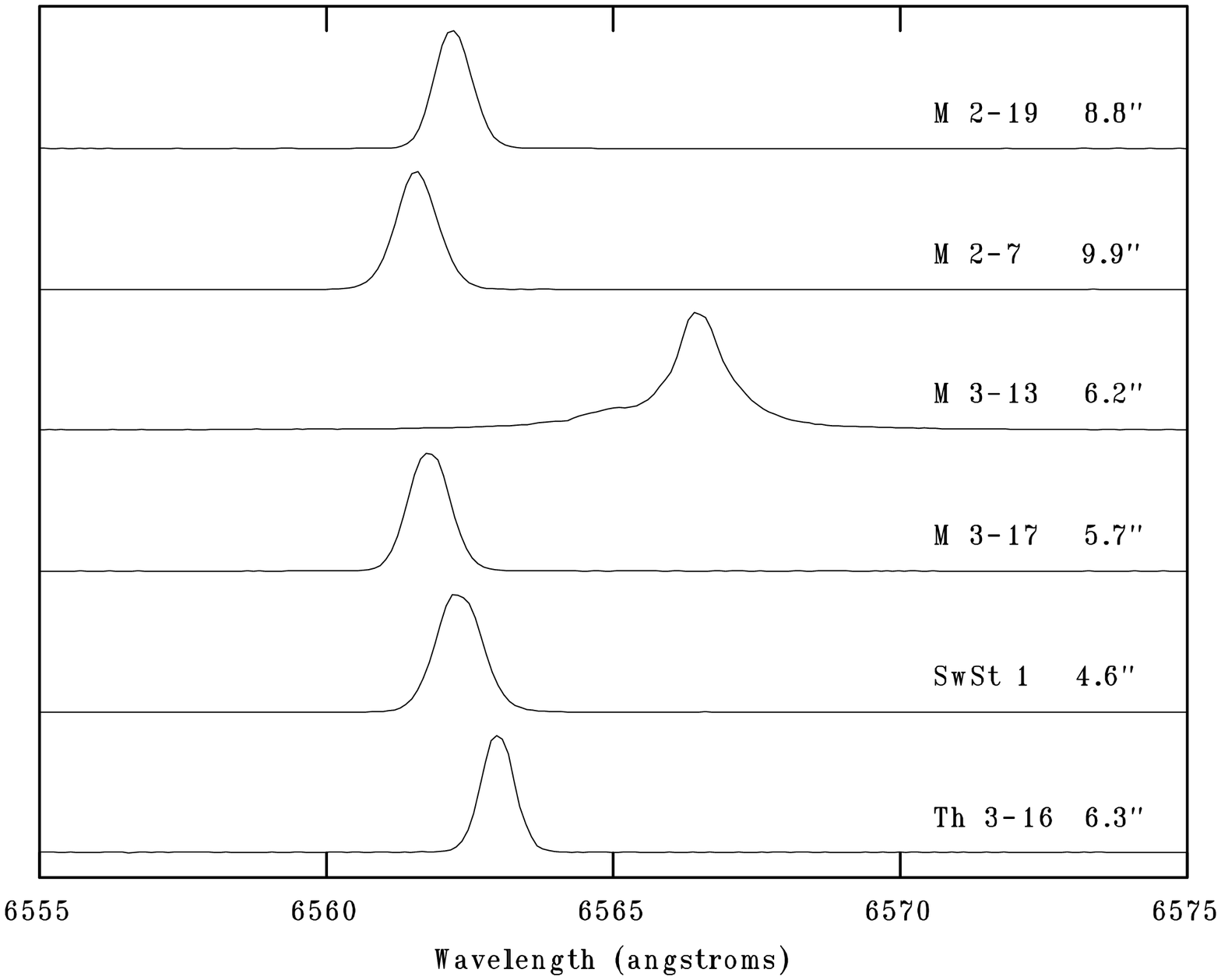}
\end{center}
\caption{As in Fig. \ref{fig_gallery_weak1}, we present the line profiles for the sample of planetary nebulae with weak [\ion{O}{3}]$\lambda$5007 lines.  Note the very wide wings on the profiles for M 3-13.      \label{fig_gallery_weak2}}
\end{figure*}

\begin{figure*}
\begin{center}
\includegraphics[height=\columnwidth,angle=90]{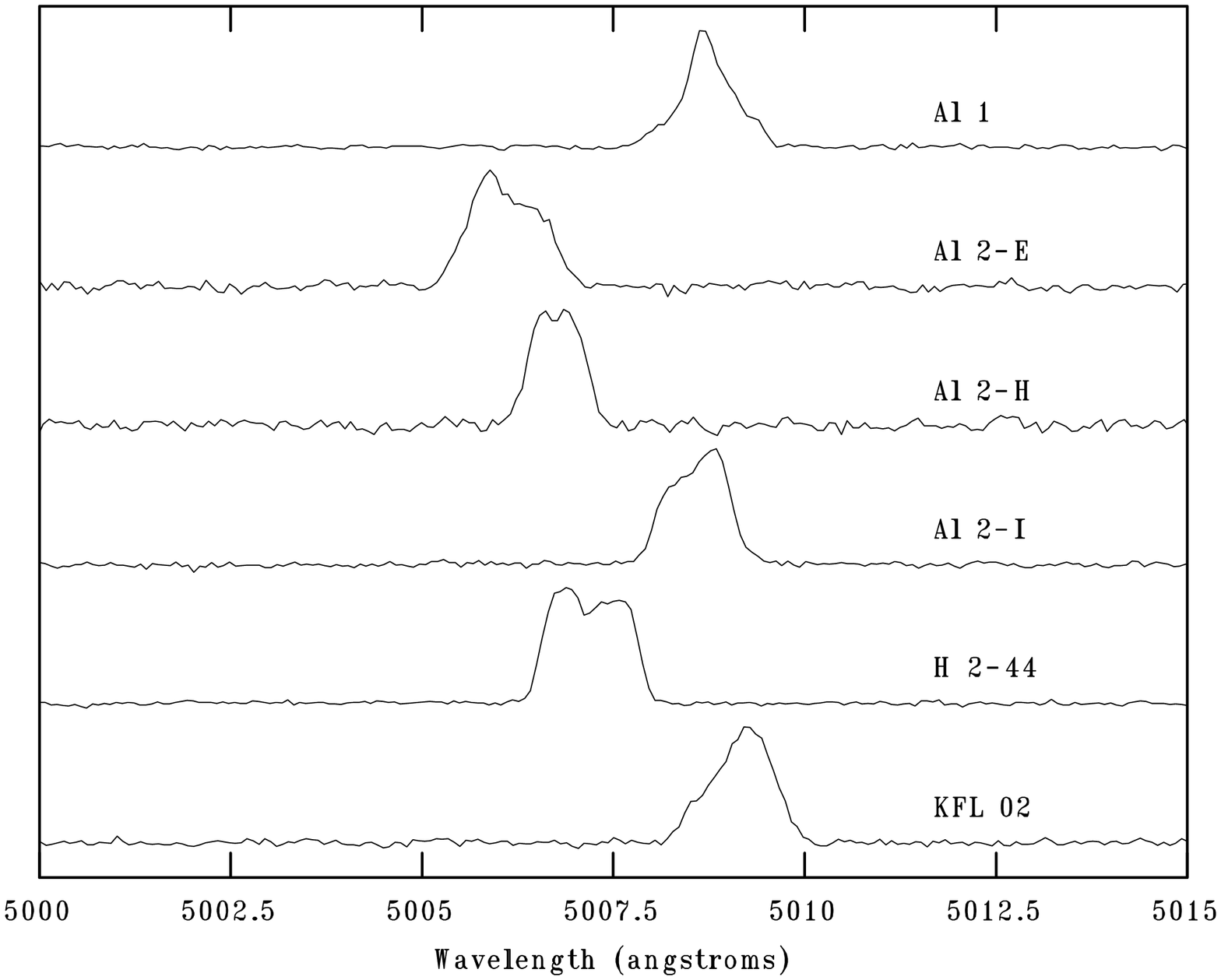}
\includegraphics[height=\columnwidth,angle=90]{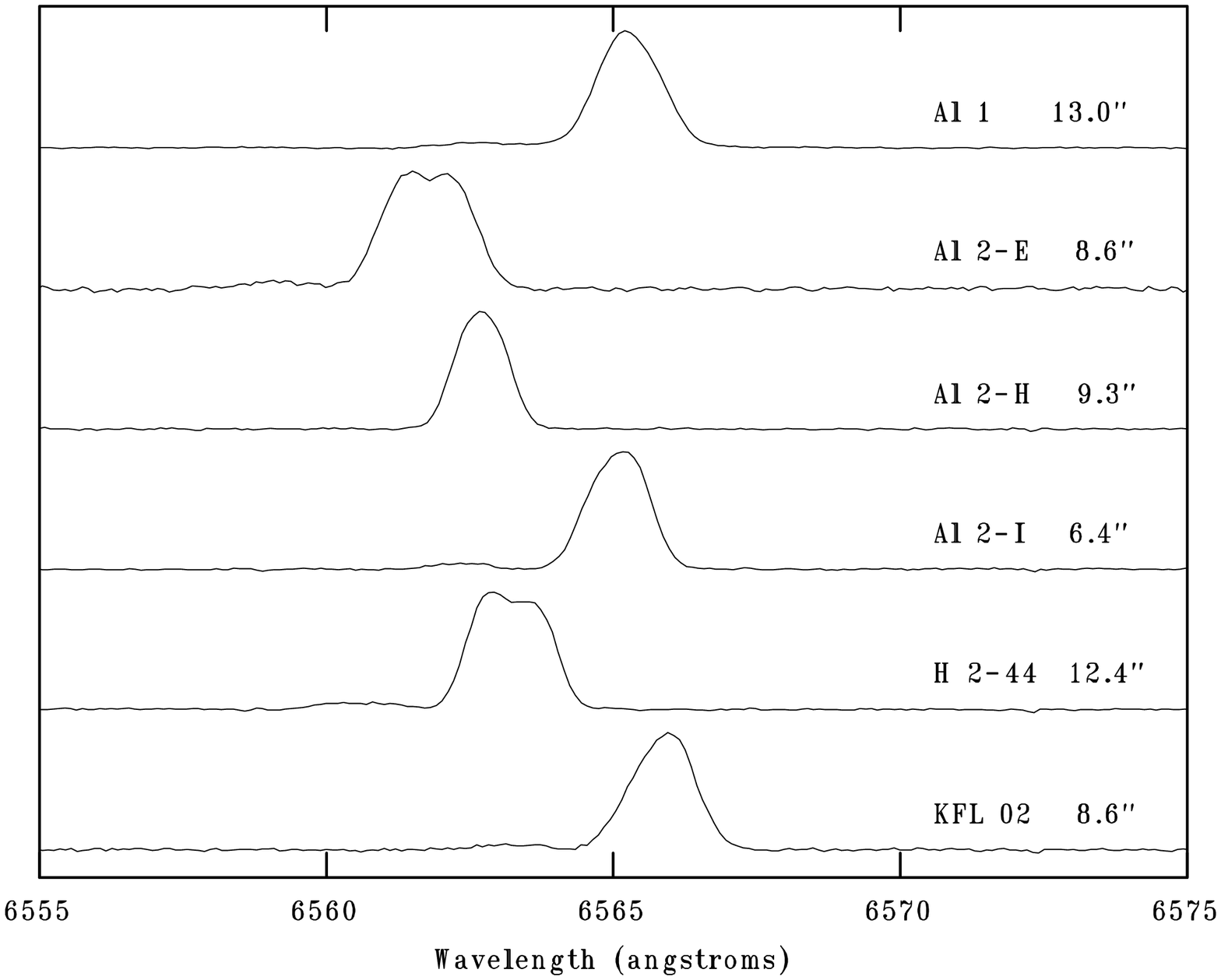} 
\includegraphics[height=\columnwidth,angle=90]{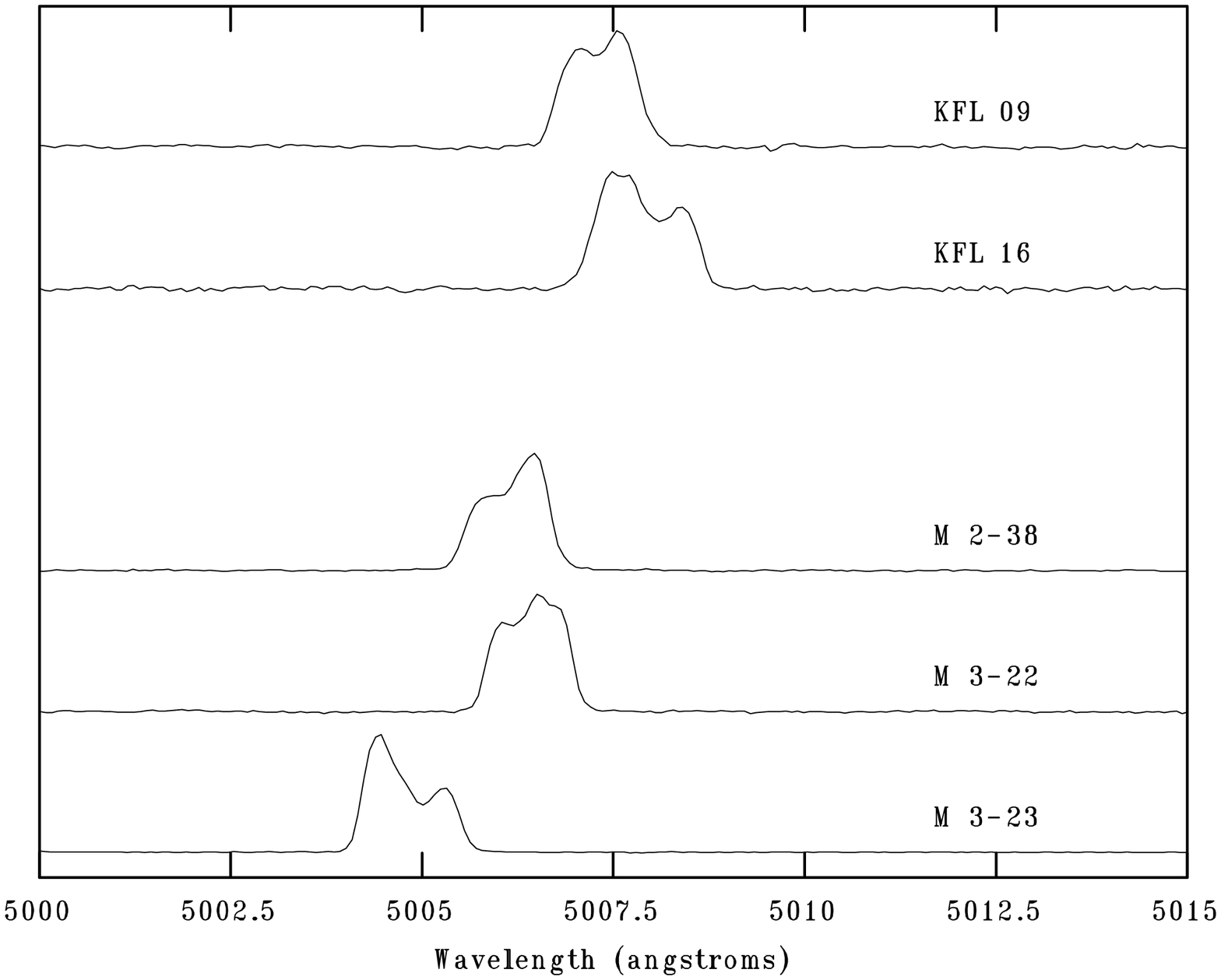}
\includegraphics[height=\columnwidth,angle=90]{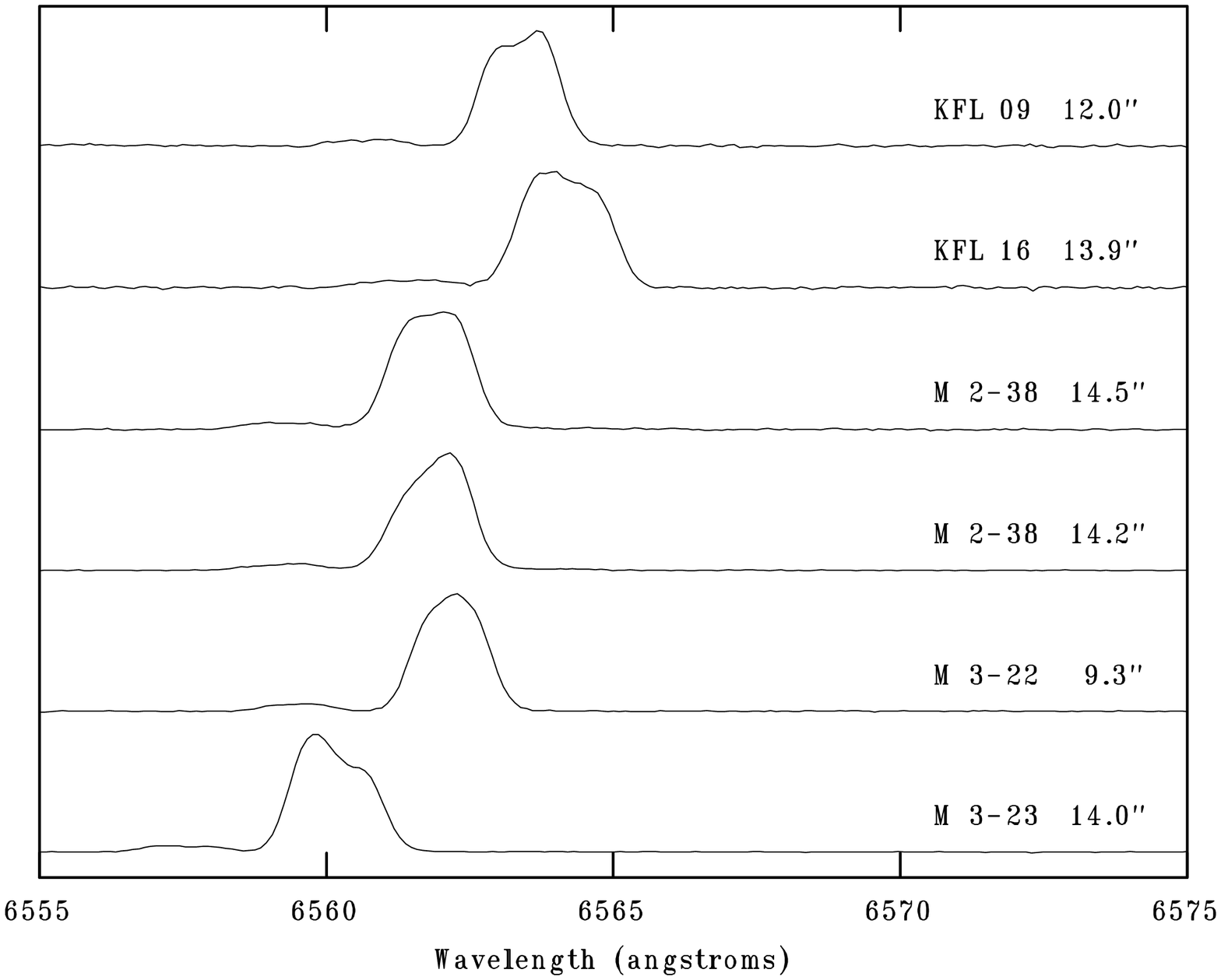}
\end{center}
\caption{As in Fig. \ref{fig_gallery_weak1}, we present the line profiles for the sample of planetary nebulae with strong \ion{He}{2} 4686 lines.  \label{fig_gallery_strong1}}
\end{figure*}

\begin{figure*}
\begin{center}
\includegraphics[height=\columnwidth,angle=90]{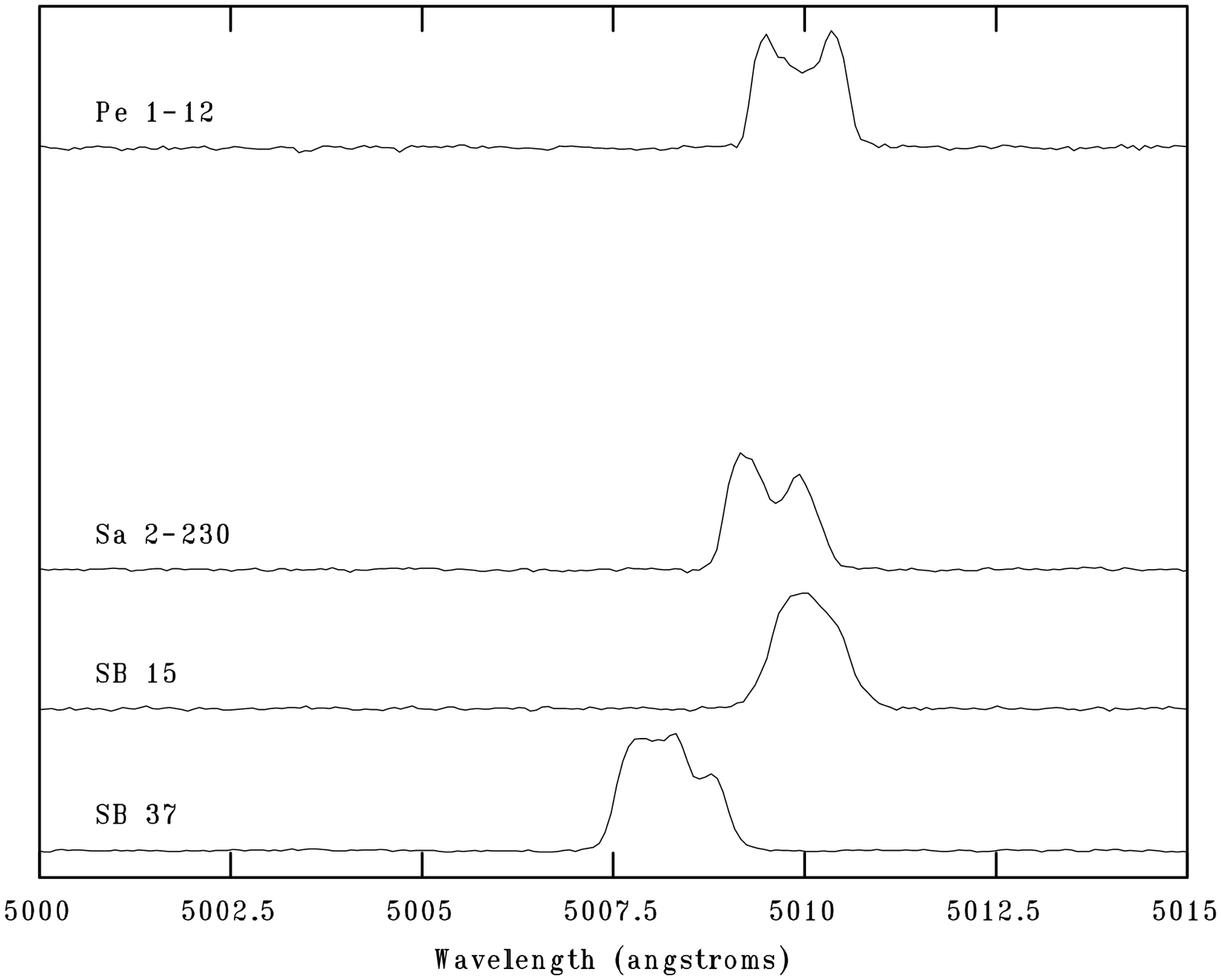}
\includegraphics[height=\columnwidth,angle=90]{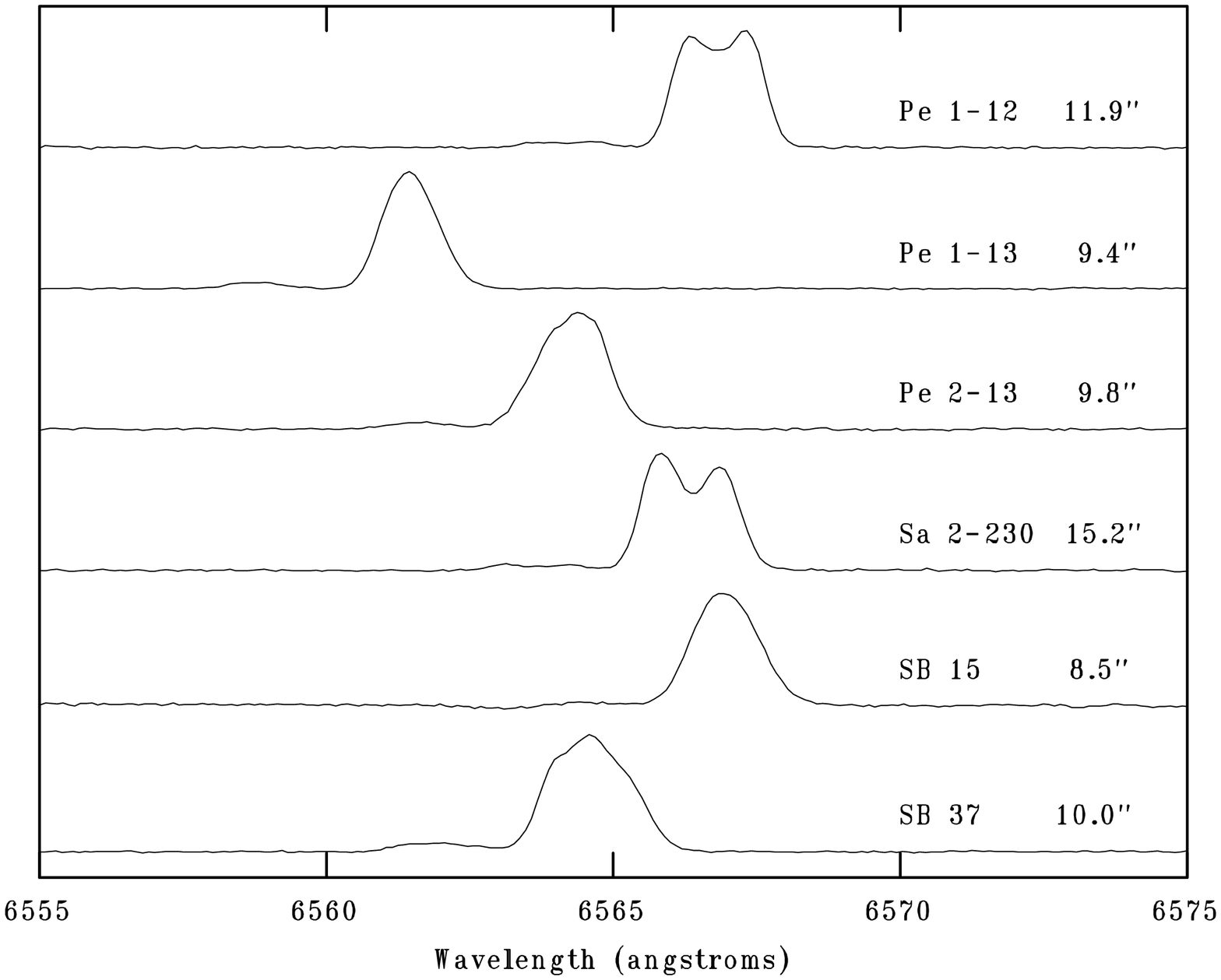} 
\includegraphics[height=\columnwidth,angle=90]{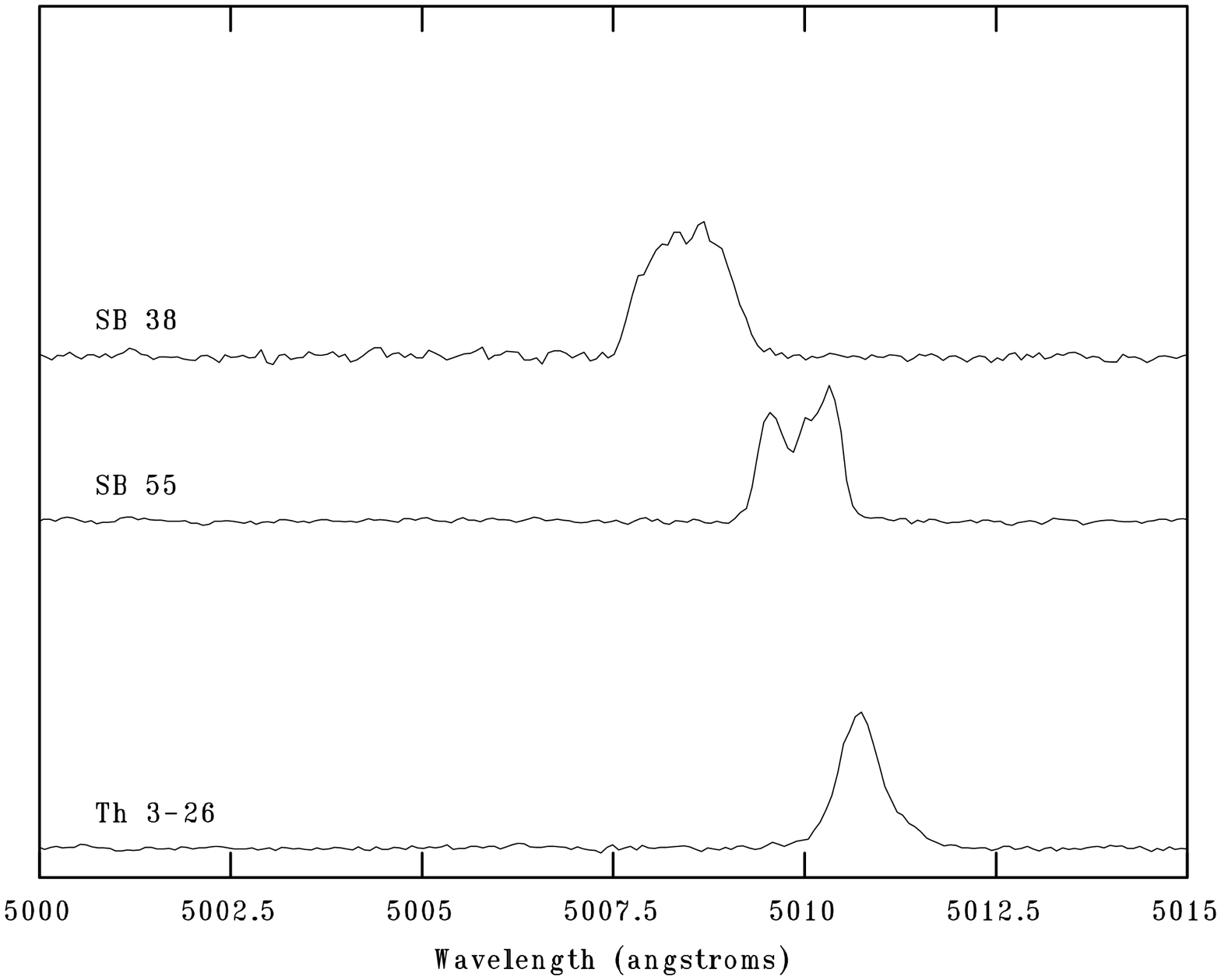}
\includegraphics[height=\columnwidth,angle=90]{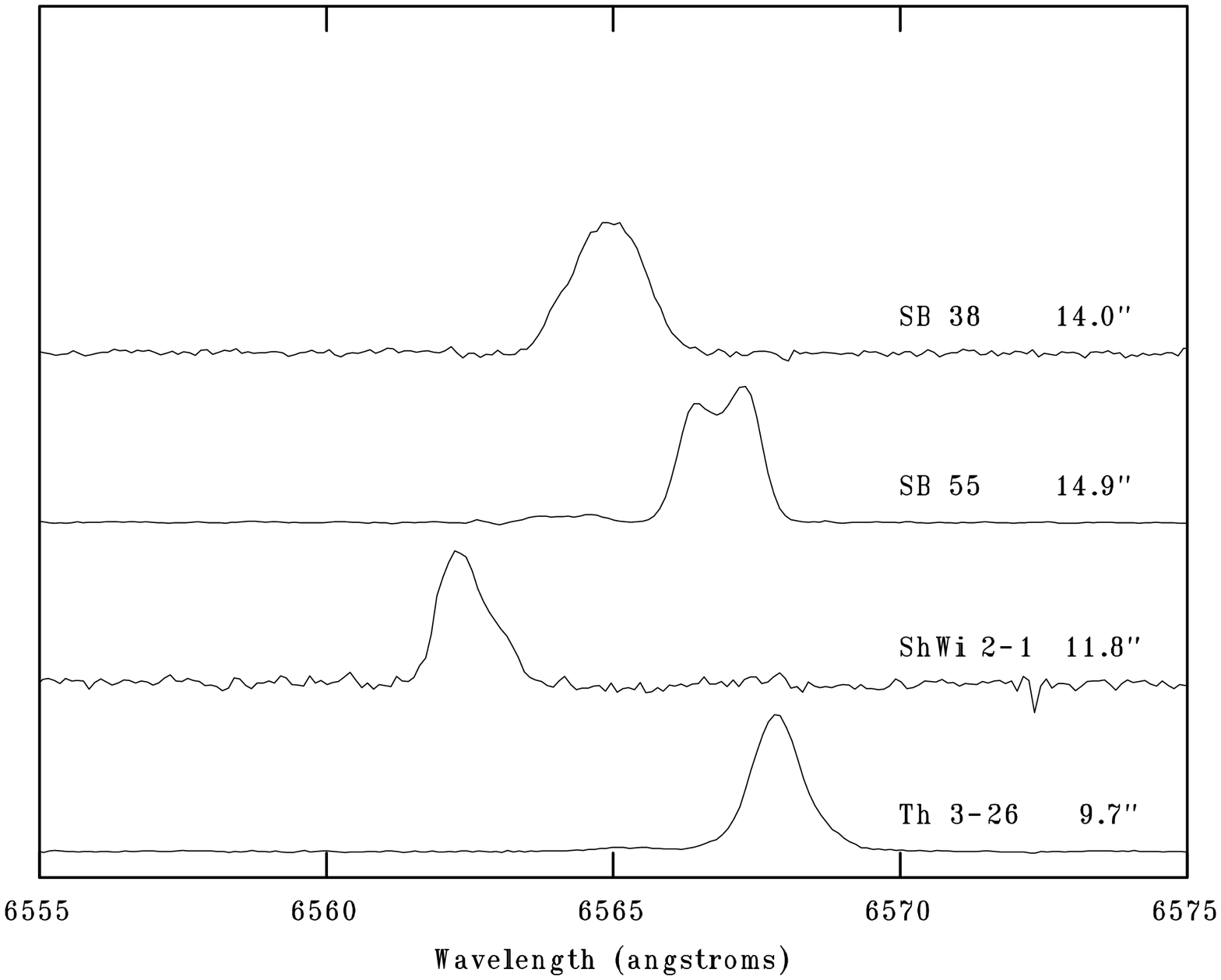}
\end{center}
\caption{As in Fig. \ref{fig_gallery_weak1}, we present the line profiles for the sample of planetary nebulae with strong \ion{He}{2} 4686 lines.  \label{fig_gallery_strong2}}
\end{figure*}

\begin{figure*}
\begin{center}
\includegraphics[scale=0.9,angle=0]{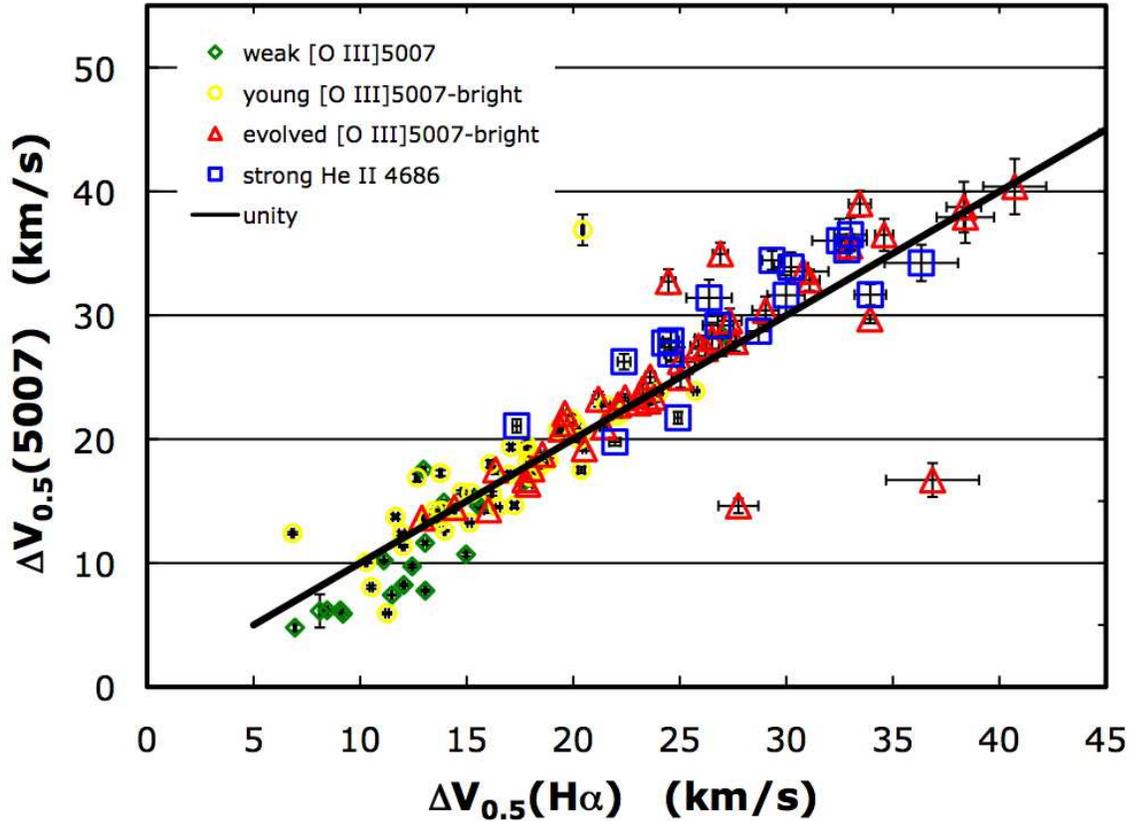}
\end{center}
\caption{The intrinsic line width in [\ion{O}{3}]$\lambda$5007 is plotted as a function of the intrinsic line width in H$\alpha$ for the two samples presented here as well as the data from \citet{richeretal2008}.  The intrinsic line widths are corrected for the instrumental resolution, fine structure broadening, and thermal broadening.  The solid line indicates the locus of identical line widths in the two emission lines.  The weak [\ion{O}{3}] $\lambda 5007$ sample extends the results of \citet{richeretal2008} to lower line widths while the strong \ion{He}{2}~$\lambda 4686$ sample is well mixed with their data.  The error bars appear to fill many of the symbols for the objects in the samples with weak [\ion{O}{3}] $\lambda 5007$ and \ion{He}{2}~$\lambda 6560$ absent.
\label{fig_dvo3_dvha}}
\end{figure*}

\begin{figure}
\begin{center}
\includegraphics[width=\columnwidth,angle=0]{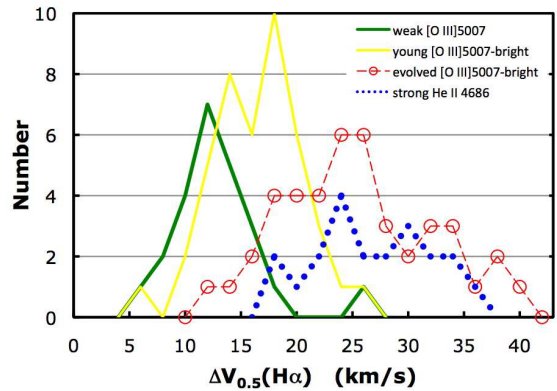}
\end{center}
\caption{%
There is a very clear increase in the line width, progressing from planetary nebulae with weak [\ion{O}{3}]$\lambda$5007 to the young and evolved [\ion{O}{3}]$\lambda$5007-bright objects, clearly demonstrating that the nebular shells are accelerated throughout these phases.  From the perspective of line widths, these three phases are statistically distinct.  Planetary nebulae in the most evolved phase, with strong \ion{He}{2}\,4686 line intensities, have line widths very similar to the evolved [\ion{O}{3}]$\lambda$5007-bright planetary nebulae.   M3-13 is the \lq\lq anomalous" object with weak [\ion{O}{3}]$\lambda$5007 and a line width of 26\,km/s (see Fig. \ref{fig_gallery_weak2}).  We correct the line widths from \citet{richeretal2008} for thermal broadening using the observed electron temperature, instead of a constant value of $10^4$\,K.  As a result, our histograms for those data differ.
\label{fig_hist_lw}}
\end{figure}

\begin{figure}
\begin{center}
\includegraphics[width=\columnwidth,angle=0]{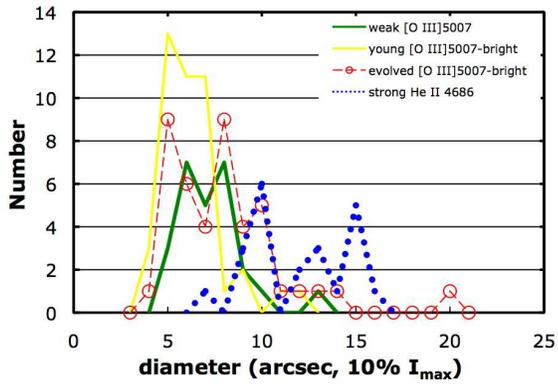}
\end{center}
\caption{%
There is a clear increase in the diameters for the planetary nebulae in the young and evolved [\ion{O}{3}]$\lambda$5007-bright and strong \ion{He}{2}\,4686 groups.  The least evolved planetary nebulae, with with weak [\ion{O}{3}]$\lambda$5007, break this evolutionary sequence.    
\label{fig_hist_d10}}
\end{figure}

\begin{figure*}
\begin{center}
\includegraphics[scale=0.9,angle=0]{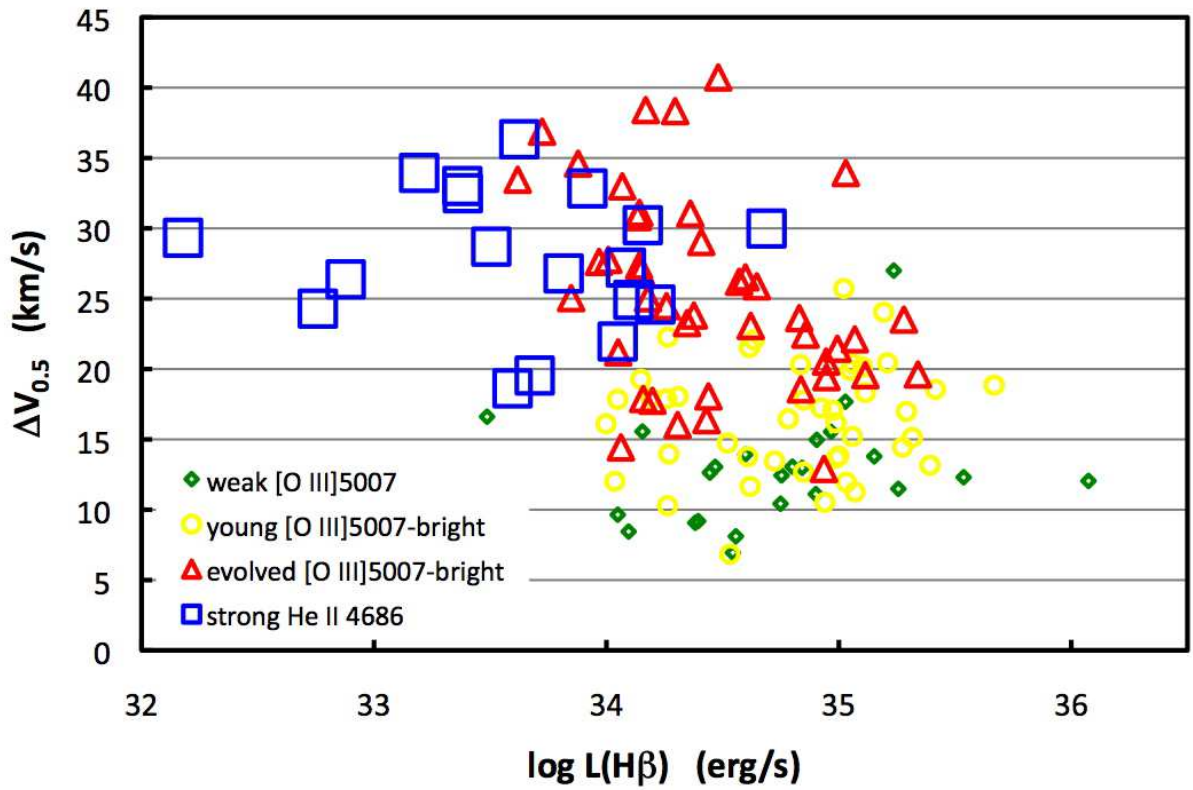}
\end{center}
\caption{The line widths are plotted as a function of the H$\beta$ luminosity for the four samples of Bulge planetary nebulae.  A variety of effects are responsible for mixing the objects from different evolutionary stages.
\label{fig_dv_hb}}
\end{figure*}


\begin{thebibliography}{}

\bibitem[Aller (1956)]{aller1956} Aller, L. H. 1956, Gaseous Nebulae (Wiley: New York)

\bibitem[Aller \& Keyes (1987)]{allerkeyes1987} Aller, L. H., \& Keyes, C. D. 1987, \apjs, 65, 405

\bibitem[Breitschwerdt \& Kahn (1990)]{breitschwerdtkahn1990} Breitschwerdt, D., \& Kahn, F. D. 1990, \mnras, 244, 521

%

\bibitem[Chu et al. (1984)]{chuetal1984} Chu, Y.-H., Kwitter, K. B., Kaler, J. B., \& Jacoby, G. H. 1984, \pasp, 96, 598

\bibitem[Cuisinier et al. (1996)]{cuisinieretal1996} Cuisinier, F., Acker, A., \& K\"oppen, J. 1996, \aap, 307, 215

\bibitem[Cuisinier et al. (2000)]{cuisinieretal2000} Cuisinier, F., Maciel, W. J., K\"oppen, J., Acker, A., \& Stenholm, B. 2000, A\&A, 353, 543

\bibitem[Dopita et al. (1985)]{dopitaetal1985} Dopita, M. A., Ford, H. C., Lawrence, C. J., \& Webster, B. L. 1985, \apj, 296, 390

\bibitem[Dopita \& Meatheringham (1990)]{dopitaetal1990} Dopita, M. A., \& Meatheringham, S. J. 1990, \apj, 357, 140

\bibitem[Dopita et al. (1988)]{dopitaetal1988} Dopita, M. A., Meatheringham, S. J., Webster, B. L., \& Ford, H. C. 1988, ApJ, 327, 639

\bibitem[Dudziak et al. (2000)]{dudziaketal2000} Dudziak, G., P\'equignot, D., Zijlstraa, A. A., \& Walsh, J. R. 2000, A\&A, 363, 717

\bibitem[Escudero \& Costa (2001)]{escuderocosta2001} Escudero, A. V., \& Costa, R. D. D. 2001, A\&A, 380, 300

\bibitem[Escudero et al. (2004)]{escuderoetal2004} Escudero, A. V., Costa, R. D. D., \& Maciel, W. J. 2004, A\&A, 414, 211

\bibitem[Exter et al. (2004)]{exteretal2004} Exter, K. M., Barlow, M. J., \& Walton, N. A. 2004, MNRAS, 349, 1291

\bibitem[Garc\'\i a-D\'\i az et al. (2008a)]{garciadiazetal2008a} Garc\'\i a-D\'\i az, Ma. T., Henney, W. J., L\'opez, J. A., \& Doi, T. 2008a, \rmxaa, 44, 181

\bibitem[Garc\'\i a-D\'\i az et al. (2008b)]{garciadiazetal2008b} Garc\'\i a-D\'\i az, Ma. T., L\'opez, J. A., Garc\'\i a-Segura, G., Richer, M. G., \& Steffen, W. 2008b, \apj, 676, 402

\bibitem[Garc\'\i a-Segura et al. (2006)]{garciaseguraetal2006} Garc\'\i a-Segura, G., L\'opez, J. A., Steffen, W., Meaburn, J., \& Manchado, A. 2006, \apjl, 646, 61

%

%

\bibitem[Gon\c calves et al. (2007)]{goncalvesetal2007} Gon\c calves, D. R., Magrini, L., Leisy, P., \& Corradi, R. L. M. 2007, MNRAS, 375, 715

\bibitem[G\'orny et al. (2004)]{gornyetal2004} G\'orny, S. K., Stasi\'nska, G., Escudero, A. V., \& Costa, R. D. D. 2004, A\&A, 427, 231

\bibitem[Groenewegen et al. (2009)]{groenewegenetal2009} Groenewegen, M. A. T., Sloan, G. C., Soszy\'nsky, I., \& Petersen, E. A. 2009, \aap, accepted; also http://xxx.lanl.gov/abs/0908.3087

\bibitem[Gurzadyan (1988)]{gurzadyan1988} Gurzadyan, G. A. 1988, ApSS, 149, 343

\bibitem[Gurzadyan (1997)]{gurzadyan1997} Gurzadyan, G. A. 1997, The Physics and Dynamics of Planetary Nebulae (Springer-Verlag: Berlin)

\bibitem[Heap (1993)]{heap1993} Heap, S. R., in IAU Symp. 155: Planetary Nebulae, eds. R. Weinberger \& A. Acker (Reidel Publishing: Dordrecht: the Netherlands), 23

\bibitem[Jacoby \& Ciardullo (1999)]{jacobyciardullo1999} Jacoby, G. H., \& Ciardullo, R. 1999, ApJ, 515, 169

\bibitem[Kahn \& West (1985)]{kahnwest1985} Kahn, F. D., \& West, K. A. 1985, \mnras, 212, 837

\bibitem[Kahn \& Breitschwerdt (1990)]{kahnbreitschwerdt1990} Kahn, F. D., \& Breitschwerdt, D. 1990, \mnras, 242, 505

\bibitem[Kwok (1982)]{kwok1982} Kwok, S. 1982, \apj, 258,280

\bibitem[Kwok et al. (1978)]{kwoketal1978} Kwok, S., Purton, C. R., \& Fitzgerald, P. M. 1978, \apjl, 219, 125

\bibitem[Lang (1980)]{lang1980} Lang, K. R. 1980, Astrophysical Formulae (Springer-Verlag: Berlin, Heidelberg)

\bibitem[Lewis (1991)]{lewis1991} Lewis, B. M. 1991, \aj, 101, 254

%

\bibitem[Marshall et al. (2004)]{marshalletal2004} Marshall, J. R., van Loon, J. Th., Matsuura, M., Wood, P. R., Zijlstra, A. A., \& Whitelock, P. A. 2004, \mnras, 355, 1348

\bibitem[Marten \& Sch\"onberner (1991)]{martenschonberner1991} Marten, H., \& 
Sch\"onberner, D. 1991, \aap, 248, 590

\bibitem [Massey et al. (1992)]{masseyetal1992} Massey, P., Valdes, F., \& Barnes, J. 1992, A User's Guide to Reducing Slit Spectra with IRAF, IRAF User Guide, Vol. 2B (Tucson: National Optical Astronomy Observatory)

\bibitem[Mattsson et al. (2008)]{mattssonetal2008} Mattsson, L., Wahlin, R., H\"ofner, S., \& Eriksson, K. 2008, \aap, 484, L5

\bibitem [McCall et al (1985)]{mccalletal1985} McCall, M. L., Rybski, P. M., \& Shields, G. A. 1985, ApJS, 57, 1

\bibitem[Meaburn (1970)]{meaburn1970} Meaburn, J. 1970, Nature, 228, 44

\bibitem[Meaburn et al. (1984)]{meaburnetal1984} Meaburn, J., Blundell, B.,
Carling, R., Gregory, D. F., Keir, D., \& Wynne, C. G. 1984, \mnras, 210, 463

\bibitem[Meaburn et al. (2003)]{meaburnetal2003} Meaburn, J., L\'opez, J. A., Guti\'errez, L., Quiroz, F., Murillo, J. M., Vald\'ez, J., \& Pedrayes, M. 2003, \rmxaa, 39, 185

\bibitem[Medina et al. (2006)]{medinaetal2006} Medina, S., Pe\~na, M., Morisset, C., \& Stasi\'nska, G. 2006, \rmxaa, 42, 53

\bibitem[Mellema (1994)]{mellema1994} Mellema, G. 1994, \aap, 290, 915

\bibitem[M\'endez et al. (2005)]{mendezetal2005} M\'endez, R. H., Thomas, D., Saglia, R. P., Maraston, C., Kudritski, R. P., \& Bender, R. 2005, ApJ, 627, 767

%

\bibitem[Perinotto et al. (2004)]{perinottoetal2004} Perinotto, M., Sch\"onberner, D., Steffen, M., \& Calonaci, C. 2004, \aap, 414, 993

\bibitem[Ramstedt et al. (2006)]{ramstedtetal2006} Ramstedt, S., Sch\"oier, F. L., Olofsson, H., \& Lundgren, A. A. 2006, \aap, 454, L103

\bibitem[Ratag et al. (1997)]{ratagetal1997} Ratag, M. A., Pottasch, S. R., Dennefeld, M., \& Menzies, J. 1997, A\&AS, 126, 297

\bibitem[Reid \& Parker (2010)]{reidparker2010} Reid, W. A., \& Parker, Q. A. 2010, PASA, in press; also http://xxx.lanl.gov/abs/0911.3689

\bibitem[Richer et al. (2009)]{richeretal2009} Richer, M. G., B\'aez, S.-H., L\'opez, J. A., Riesgo, H., \& Garc\'\i a-D\'\i az, Ma. T. 2009, \rmxaa, 45, 239

\bibitem[Richer et al. (2008)]{richeretal2008} Richer, M. G., L\'opez, J. A., Pereyra, M., Riesgo, H., Garc\'\i a-D\'\i az, M. T., \& B\'aez, S.-H. 2008, \apj, 689, 203

%

\bibitem[Richer \& McCall (2008)]{richermccall2008} Richer, M. G., \& McCall, M. L. 2008, \apj, 684, 1190

\bibitem[Richer et al. (1999)]{richeretal1999} Richer, M. G., Stasi\'nska, G., \& McCall, M. L. 1999, A\&AS, 135, 203

\bibitem[Richer et al. (2010)]{richeretal2010} Richer, M. G., 
%
et al. 2010, \rmxaa, submitted

\bibitem[Roth et al. (2004)]{rothetal2004} Roth, M. M., Becker, T., Kelz, A., \& Schmoll, J. 2004, ApJ, 603, 531

%

\bibitem[Sahu et al. (2006)]{sahuetal2006} Sahu, K. C., 
%
et al. 2006, Nature, 443, 534

\bibitem[Schmidt-Voigt \& K\"oppen (1987a)]{schmidtvoigtkoppen1987a} Schmidt-Voigt, M., \& K\"oppen, J. 1987a, \aap, 174, 211

\bibitem[Schmidt-Voigt \& K\"oppen (1987b)]{schmidtvoigtkoppen1987b} Schmidt-Voigt, M., \& K\"oppen, J. 1987b, \aap, 174, 223

\bibitem[Sch\"onberner et al. (2005a)]{schonberneretal2005} Sch\"onberner, D., Jacob, R., \& Steffen, M. 2005a, A\&A, 441, 573

\bibitem[Sch\"onberner et al. (2005b)]{schonberneretal2005b} Sch\"onberner, D., Jacob, R., \& Steffen, M., \& Roth, M. M. 2005b, in Planetary Nebulae as Astronomical Tools, eds. R. Szczerba, G. Stasi\'nska, \& S. G\'orny, AIP Conference Proceedings, 804, 269

\bibitem[Sch\"onberner et al. (2007)]{schonberneretal2007} Sch\"onberner, D., Jacob, R., Steffen, M., \& Sandin, C. 2007, \aap, 473, 467

\bibitem[Tylenda et al. (1994)]{tylendaetal1994} Tylenda, R., Stasi\'nska, G., Acker, A., \& Stenholm, B. 1994, \aaps, 106, 559

%
%
%

\bibitem[Villaver et al. (2002)]{villaveretal2002} Villaver, E., Manchado, A., \& Garc\'\i a-Segura, G. 2002, \apj, 581, 1204

\bibitem[Wachter et al. (2008)]{wachteretal2008} Wachter, A., Winters, J. M., Schr\"oder, K.-P., \& Sedlmayr, E. 2008, \aap, 486, 497

\bibitem[Wall \& Jenkins (2003)]{walljenkins2003} Wall, J. V., \& Jenkins, C. R. 2003, Practical Statistics for Astronomers (Cambridge University Press: Cambridge, U.K.)

\bibitem[Walsh et al. (1999)]{walshetal1999} Walsh, J. R., Walton, N. A., Jacoby, G. H., \& Peletier, R. F. 1999, A\&A, 346, 753

\bibitem[Webster (1988)]{webster1988} Webster, B. L. 1988, \mnras, 230, 377

\bibitem[Wilson (1950)]{wilson1950} Wilson, O. C. 1950, \apj, 111, 279

\bibitem[Wood et al. (1992)]{woodetal1992} Wood, P. R., Whiteoak, J. B., Hughes, S. M. G., Bessell, M. S., Gardner, F. F., \& Hyland, A. R. 1992, \apj, 397, 552

\bibitem[Zijlstraa et al. (2006)]{zijlstraaetal2006} Zijlstraa, A. A., Gesicki, K., Walsh, J. R., P\'equignot, D., van Hoof, P. A. M., \& Minniti, D. 2006, \mnras, 369, 875

\bibitem[Zoccali et al. (2008)]{zoccalietal2008} Zoccali, M., Hill, V., Lecureur, A., Barbuy, B., Renzini, A., Minniti, D., G\'omez, A., \& Ortolani, S. 2008, \aap, 486, 177

\end{thebibliography}
\end{document}